\documentclass[10pt]{article}
\usepackage{amssymb}
\usepackage{amsfonts}
\usepackage{bbold}

\usepackage{geometry}

\usepackage{amsmath}
\usepackage{amsthm}
\usepackage{mathrsfs} 
\usepackage{graphicx}
\usepackage{mathtools}
\usepackage{enumitem}
\usepackage{caption}
\usepackage[nameinlink]{cleveref}
\numberwithin{equation}{section}
\usepackage{xcolor}

\usepackage{yfonts}
\usepackage{amssymb}
\usepackage{amsmath}
\theoremstyle{theorem}
\newtheorem{theorem}{Theorem}[section]

\newtheorem{proposition}[theorem]{Proposition}

\newtheorem{definition}[theorem]{Definition}
\newtheorem{remark}[theorem]{Remark}

\newtheorem{example}[theorem]{Example}

\def\spa{\hskip -3pt}

\def\bearray{\begin{eqnarray}}
\def\earray{\end{eqnarray}}
\def\beq{\begin{equation}}
\def\eeq{\end{equation}}

\def\b0{{\bf 0}}

\def\cA{{\cal A}}

\newsymbol\bt 1202             

\def\bC{{\mathbb C}}           

\def\bN{{\mathbb N}}

\def\bR{{\mathbb R}}

\def\bZ{{\mathbb Z}} 

\newsymbol\rest 1316         
\def\gA{{\mathfrak A}}       
\def\gB{{\mathfrak B}}

\begin{document} 

\par
\bigskip
\large
\noindent
{\bf The classical limit and spontaneous symmetry breaking in algebraic quantum theory}
\bigskip
\par
\rm
\normalsize

\noindent {\bf Christiaan J.F.  van de Ven}\\
\par

\noindent 
Max Planck Institute for Mathematics in the Sciences,\\
Inselstra\ss e 22, 04103 Leipzig, Germany.\\
Email: vandeven@mis.mpg.de 

\par

\rm\small

\rm\normalsize

\par
\bigskip

\noindent
\small
{\bf Abstract.} In this paper an overview of  some recent developments on the classical limit and spontaneous symmetry breaking (SSB) in algebraic quantum theory is given.  In such works, based on the theory of $C^*$-algebras, the concept of the classical limit has been formalized in a complete algebraic manner. Additionally, since this setting allows for commutative as well as non-commutative $C^*$-algebras, and hence for classical and quantum theories, it provides an excellent framework to study SBB as an emergent phenomenon when transitioning from the quantum to the classical world by turning off a semi-classical parameter. We summarize the main results and show that this algebraic approach sheds new light on the connection between the classical and the quantum realm, where particular emphasis is placed on the role of SSB in Theory versus Nature. To this end a detailed analysis is carried out and illustrated with three different physical models: Schr\"{o}dinger operators, mean-field quantum spin systems and the Bose-Hubbard model.
\normalsize
\newpage
\tableofcontents

\section{Introduction}\label{introduction}

For decades the natural phenomenon of spontaneous symmetry breaking (SSB) has been a topic of great interest in mathematical physics and theoretical physics. It forms the basis of many physical phenomena, including phase transitions in condensed-matter systems, superconductivity of metals and it is the origin of particle masses in the standard model, described by the Higgs mechanism \cite{Ruelle,Wez2}. Intensive studies have led to important results and insights concerning symmetry and its possible breakdown in a various number of physical models. 

The general and common concept behind spontaneous symmetry breaking, originating in the field of condensed matter physics where one typically considers the limit of large particle numbers, often also called thermodynamic limit, is based on the idea that if a collection of quantum particles becomes larger, the symmetry of the system as a whole becomes more unstable against small perturbations \cite{Wez2,VGRL18}. A similar statement can be made for quantum systems in their classical limit, where sensitivity against small perturbations now should be understood to hold in the relevant semi-classical regime, meaning that a certain parameter (e.g. $\hbar$) approaches zero\footnote{Mathematically, this parameter is an element of the base space corresponding to a $C^*$-bundle (see Definition \ref{def:continuous algebra bundle}). In the context of the classical limit,  the zero-limit of this parameter corresponds to a classical theory, encoded by a commutative $C^*$-algebra. We stress that the precise interpretation of this parameter depends on the physical situation.} at fixed system size \cite{Lan17,MorVen2}.

Showing the occurrence of SSB in a certain particle system can be done at various levels of rigour.  Mathematically, there are some differences between the mathematical physics approach (used in this paper) to SSB in finite quantum systems and the standard methods used in theoretical and condensed matter physics. In the latter approaches, the main concept that the relevant system becomes sensitive to small perturbations is generally taken into account by adding a so-called infinitesimal symmetry breaking field term \cite{KoTa,Wez1,Wez2}. Consequently, one aims to show that the limit of large particle numbers (or thermodynamic) limit becomes “singular'', at least at the level of states, e.g. the ground state. If this happens, one says that the symmetry of the limiting system is spontaneously broken. To get an idea what this means let us consider the quantum Curie-Weiss Hamiltonian $H_{1/N}^{CW}$ (see $\S$\ref{applicationsCW} for a more general discussion and details regarding this model), i.e.
\begin{align}
H^{CW}_{1/N} =-\frac{J}{2N} \sum_{i,j=1}^N \sigma_3(i)\sigma_3(j) -B \sum_{j=1}^N \sigma_1(j),
\end{align}
where $B\in (0,1)$ denotes the magnetic field and $J$ a coupling constant that can be chosen to be one. The symmetry-breaking term is typically taken to be
\begin{align}
\delta_{1/N}^{CW}=\epsilon\sum_{x=1}^N\sigma_3(x).
\end{align}
In this approach originating with the ideas of Bogoliubov, one argues that the correct order of the limits should be $\lim\epsilon\to 0 \lim N\to\infty$ \cite{VGRL18,Wez2}, which gives SSB by one of the two pure classical ground states on the limit algebra $C(B^3)$, with $B^3$ the closed unit ball in $\mathbb{R}^3$, where the sign of $\epsilon$ determines the direction of symmetry breaking.\footnote{We refer to $\S$\ref{applicationsCW} for details on these states and the construction of this algebra.}  In contrast, the opposite order $\lim N\to\infty \lim \epsilon\to 0$ gives a  symmetric but mixed  ground state on the limit algebra. \footnote{This mixture is precisely the one determined by Theorem \ref{mainsecond2}.} It is then said that the symmetry is broken spontaneously if there is a difference in the order of the limits, as exactly happens in this example. In this sense, if SSB occurs, the limit $N\to\infty$ can indeed be seen as singular \cite{Batterman,Berry,Wez2}. However, this approach to SSB is not common in mathematical physics and furthermore challenged in the philosophical literature of physics \cite{Butterfield}.

Instead, this paper is based on a definition of SSB that is standard in mathematical physics.  It stands on an algebraic formulation of symmetries and ground states \cite{BR1,BR2,Lan17} carefully explained in Section \ref{symminalgquant}. This approach equally applies to finite and infinite systems, and to classical and quantum systems, namely that the ground state, suitably defined of a system with $G$-invariant dynamics (where $G$ is some group, typically a discrete group or a Lie group) is either pure  but not $G$-invariant, or $G$-invariant but mixed.\footnote{Strictly speaking one should consider extremal ground states, which in many cases of physical interest
turn out to be exactly the pure ground states \cite{Lan17} (see also Section \ref{symminalgquant}).}

\begin{remark}
{\em  It may perhaps seem more natural to only require that the ground state fails to be $G$-invariant. However, since in the $C^*$-algebraic formalism ground states that are not necessarily pure are taken into account as well, this gives the possibility of forming $G$-invariant mixtures of non-invariant states that lose the purity properties one expects  ground states to have. A similar statement holds for equilibrium states, where “pure'' is replaced by “primary'', which corresponds to a mathematical property of pure thermodynamic phases \cite{BR1,BR2,Lan17}.
}
\hfill$\blacksquare$
\end{remark}
  Accordingly, what is singular about the thermodynamic limit of systems with SSB is the fact that the exact  pure ground state of a finite quantum system converges to a mixed state on the limit system, explained in detail in Section \ref{Applications} (see also \cite{LMV,MorVen2,Ven2020}).  In this algebraic approach the general physical idea that spontaneous symmetry breaking should be related to instability and sensitivity of the system against small perturbations in the relevant regime (see previous discussion) is elucidated in $\S$\ref{SSBINNATURE}.

\subsection{SSB as emergent phenomenon}
Historically, the concept of spontaneous symmetry breaking first emerged in condensed matter physics. The prototype case is the antiferromagnetic quantum Heisenberg model.
\begin{example}[Antiferromagnetic quantum Heisenberg chain]\label{antiferroheisenberg}
{ \em
Consider the QH Hamiltonian \cite{KoTa1993,Tasaki19}
\begin{align}
H_L^{QH}=\sum_{x,y\in\Lambda_L, |x-y|=1}\mathbf{S}_x\cdot \mathbf{S}_y,
\end{align}
on a one-dimensional chain $\Lambda_L$ with even $L=|\Lambda_L|$ and we impose periodic boundary conditions. On each site $x$, $\mathbf{S}_x=(S_x^1,S_x^2,S_x^3)$ is a quantum spin operator on $\bC^{2J+1}$ ($2J$ integer):
\begin{align}
&[S_x^j,S_y^k]=i\epsilon_{jkl}S_x^l\delta_{x,y}; \\
& (\mathbf{S}_x)^2=(S_x^1)^2+(S_x^2)^2+(S_x^3)^2=J(J+1),
\end{align}
acting on the Hilbert space $\mathcal{H}\equiv\mathcal{H}(J,L)\cong \bigotimes_{x\in\Lambda_L}\bC^{2J+1}$, where $J$ denotes the angular momentum of the spin operator at site $x$. The Hamiltonian  $H_L^{QH}$ is invariant under $SU(2)$ symmetry. For finite $L$, it is a well-known fact that the ground state eigenvector is unique and therefore $SU(2)$ -invariant: no SSB occurs. Instead, it can be shown that in the limit $L\to\infty$ the ground state of the antiferromagnetic model becomes infinitely degenerate and loses its $SU(2)$ invariance, i.e. the rotation symmetry is spontaneously broken \cite{KoTa1993}.\footnote{The limit $L\to\infty$ corresponds to the thermodynamic limit (viz. Section \ref{algfor}, in particular Example \ref{spin2}). In $C^*$-algebraic language this means that the algebraic ground state defined for each $L$ admits, as $L\to\infty$, a limit ( taken w.r.t quasi-local observables) as a state on the corresponding non-commutative quasi-local $C^*$-algebra $\gB_0^l$ (cf. \eqref{B0l}). A general discussion can be found in \cite{ BR1,BR2,Lan17}.} In other words, SSB shows up as emergent phenomenon in the limit of large particle numbers.
}
\hfill$\blacksquare$
\end{example}

In quantum theory, the crucial point is that for finite systems the ground state (or the equilibrium state) a generic Hamiltonian is unique\footnote{Perhaps the physically most famous  exception to this idea is the ferromagnetic quantum Heisenberg model, where the ground state for any finite $L$ is already degenerate.} and hence invariant under whatever symmetry group $G$ it may have \cite{Lan13,VGRL18}. Hence, mathematically speaking, the possibility of SSB, in the sense of having a family of asymmetric pure ground states related by the action of $G$ (viz. Definition \ref{def:SSB} in $\S\ref{weakandssb}$), seems to be reserved for infinite systems seen as thermodynamic limit of the pertinent finite system. Analogously, in view of the classical limit (often denoted by $\hbar\to 0$, see $\S$\ref{sec:classicallimit}), an exact similar situation occurs in quantum mechanics of finite systems (which typically forbids SSB), versus classical mechanics of finitely many degrees of freedom\footnote{We refer to \cite{GV} for a discussion on classical systems with infinitely many degrees of freedom. Besides the fact that such systems allow SSB as well, they are moreover used to study (classical) phase transitions which are common for infinite systems \cite{BRW}. In this paper all classical theories we consider are finite, i.e. they are encoded by a finite-dimensional phase space.}, which allows it. 
Therefore, generally speaking spontaneous symmetry breaking can be seen as natural {\em emergent phenomenon}\footnote{We refer to \cite{Lan13} for a detailed discussion on this topic.}, meaning that is only shows up in the limit (an infinite quantum or a finite classical system) of an underlying finite quantum theory. 

In this paper we review recent developments based on the theory of $C^*$-algebras, and show that the methods used in such works shed new light on the classical limit and spontaneous symmetry breaking.
The paper is structured as follows. In Section \ref{algfor} we first introduce the general definitions (Def. \ref{def:continuous algebra bundle} in $\S$\ref{contbundleofc} and Def. \ref{def:deformationq} in $\S$\ref{deformationquantization}). Consequently, in Section \ref{symminalgquant} the concepts and definitions of ground state (Def. \ref{def:grounds} in $\S$\ref{groundsymm}) and spontaneous symmetry breaking (Def. \ref{def:SSB} in $\S$\ref{weakandssb}) are given. Finally, in Section \ref{Applications} we apply this framework to several physical models, each of a different origin. In Section \ref{Discussion} we discuss the relation between symmetry breaking in Theory versus symmetry breaking in Nature. We furthermore pose some open problems and present further research.

\section{Algebraic formalism}\label{algfor}
There exist several approaches to give a precise meaning to the limits $N\to\infty$ or $\hbar\to 0$. For Schr\"{o}dinger operators $H_\hbar$ one can not simply put $\hbar=0$ in front of the Laplacian as the resulting  operator (i.e. the potential) has nothing to do with the $\hbar$-semiclassical behaviour of the operator $H_\hbar$ itself. A similar result occurs for spin Hamiltonians $H_N$ indexed by, e.g. the number of particles $N$. The limit $N\to\infty$ of $H_N$ is even undefined! A possibility to give a precise meaning to the limits $N\to\infty$ or $\hbar\to 0$ is to introduce an algebraic framework including both quantum (i.e. $N<\infty$ or $\hbar>0$) as the limiting theory ($N=\infty$ or $\hbar=0$). This idea, dating back to Dixmier \cite{Dix} and reformulated by Kirchberg \& Wassermann \cite{Kirchberg-Wassermann}, is that both theories are reformulated in terms of a family of $C^*$-algebras  $(\gA_\hbar)_{\hbar\in I}$ which are glued together by specifying a topology on the disjoint union $\amalg_{\hbar\in I}A_\hbar$, seen as a fiber bundle over $I$ \cite{Lan17}. This topology may in fact be given rather indirectly, namely via the specification of the space of continuous sections. This framework exists under the name {\em continuous bundle of $C^*$-algebras}. 

\subsection{Continuous bundle of $C^*$-algebras}\label{contbundleofc}
Let us give the definition of a continuous bundle of $C^*$-algebras \cite[Def. C.121]{Lan17}.

\begin{definition}\label{def:continuous algebra bundle}
A  {\bf $C^*$-bundle}\footnote{Often called {\em (continuous) field of $C^*$-algebras}.} is a triple $\cA:=(I,\gA,\pi_\hbar:\gA\to \gA_\hbar)$,
 where $I$ is a locally compact Hausdorff space,
 $\gA$ is a complex $C^*$-algebra, $\{\gA_\hbar\}_{\hbar\in I}$ is a collection of $C^*$-algebras and $\pi_\hbar:\gA\to \gA_\hbar$
 is a
surjective homomorphism  of complex $C^*$-algebras for each $\hbar \in I$, such that
\begin{itemize}
\item[(i)] $\|a \|= \sup_{\hbar\in I} \|\pi_\hbar(a)\|_\hbar$,  where $\|\cdot\|$ (resp. $\|\cdot\|_\hbar$) denoting the $C^*$-norm of $\gA$ (resp. $\gA_\hbar$);
\item[(ii)] there exists an action $C_0(I) \times \gA \to \gA$ satisfying $f(\hbar)\pi_\hbar(a) = \pi_\hbar(f a)$ for any $\hbar\in I$ and $f\in C_0(I)$.\footnote{The set $C_0(I)$ denotes the space of continuous functions over $I$ vanishing at infinity.}
\end{itemize} 
A {\bf section} of the bundle is an element $\{a_{\hbar}\}_{\hbar\in I}$ of $\Pi_{\hbar\in I}\gA_\hbar$ for which there exists an $a\in \gA$
 such that $a_\hbar=\pi_\hbar(a)$ for each $\hbar\in I$. A {\bf $C^*$-bundle} $\cA$ is said to be {\bf continuous}, and its sections are called {\bf continuous sections},  if they satisfy
\begin{itemize}
\item[(iii)] for $a \in \gA$, the norm function $I \ni\hbar \mapsto \|\pi_\hbar(a)\|_\hbar$ is in $C_0(I)$. 
\end{itemize}
Cross-sections are also denoted by $\sigma$, i.e. maps $\sigma:I\to \gA_\hbar$ satisfying the above requirements.
\end{definition} 

\begin{remark}{\em Since the $\pi_\hbar$ are homomorphisms of $C^*$-algebras,  the $*$-algebra operations in $\gA$ are induced by the corresponding pointwise operations of the sections $I\ni \hbar \mapsto \pi_\hbar(a)$. It follows that $\gA$ may be identified with the space of continuous sections of the bundle, and under this identification the homomorphism $\pi_\hbar$ is just the evaluation map at $\hbar$.}
\hfill$\blacksquare$
\end{remark}

For purpose of this paper we focus on four different $C^*$-bundles, each of which with a different application to physics. These are outlined in the following examples.

\begin{example}\label{schr}
{\em
We put
\begin{align}
\gA_0^{c}&=C_0(\bR^{2n})  \ \ (\hbar=0);\\
\gA_\hbar&=B_{\infty}(L^2(\bR^n)) \ \ (\hbar>0),
\end{align}
where $C_0(\bR^{2n})$ are the continuous functions over $\bR^{2n}$ vanishing at infinity and $B_{\infty}(L^2(\bR^n))$ is the $C^*$-algebra of compact operators on $L^2(\bR^n)$. Then, $\gA_0^c$ and $\gA_\hbar$ are the fibers of a continuous bundle of $C^*$-algebras $\gA^c$ over $I=[0,1]$ \cite[Prop. II 2.6.5]{Lan98}. As a result of \cite[Prop. II. 1.2.3]{Lan98} the continuous cross-sections are given by all sequences $(a_{\hbar})_{\hbar\in I}\in\Pi_{\hbar\in I}\gA_{\hbar}$ for which $a_0\in C_0(\bR^{2n})$ and $a_{\hbar}\in \gA_{\hbar}$  and such that the sequence  $(a_{\hbar})_{\hbar\in I}$ is asymptotically equivalent to $(Q_{\hbar}^B(a_0))_{\hbar\in I}$, in the sense that
\begin{align}
\lim_{\hbar\to 0}||a_{\hbar}-Q_{\hbar}^B(a_0)||=0,
\end{align}
where $Q_\hbar^B$ denotes the Berezin quantization map defined for $f\in C_0(\bR^{2n})$ (see e.g. \cite{MorVen2}) as
\begin{align}
 Q_\hbar^B(f):= \int_{\bR^{2n}} f(q,p) |\Psi_\hbar^{(q,p)}\rangle \langle \Psi_\hbar^{(q,p)}| \frac{dq dp}{(2\pi\hbar)^n}, \label{defQB}
\end{align}
and $|\Psi_\hbar^{(q,p)}\rangle \langle \Psi_\hbar^{(q,p)}|$ denotes the orthogonal projection onto the linear space spanned by the vector $\Psi_\hbar^{(q,p)}$, which is, for given $(q,p) \in \bR^{2n}$, defined as
\begin{align}\label{vecPSI}
\Psi_\hbar^{(q,p)}(x) := \frac{e^{-\frac{i}{2}p\cdot q/\hbar} e^{ip\cdot x/\hbar} e^{-(x-q)^2/2\hbar}}{(\pi \hbar)^{n/4}} \:, \quad x \in \bR^n\:, \hbar >0.\:
\end{align}
The vector $\Psi_\hbar^{(q,p)}$ is a unit vector in $L^2(\bR^n, dx)$ also called a {\bf Schr\"{o}dinger coherent state}.
As a result, for each $f$, a particular choice of a continuous cross-section $\sigma_f$ of $\gA^c$ is given by
\begin{align}
&\sigma_f:0\mapsto f\in \gA_0^{c};\\
&\sigma_f:\hbar\mapsto Q_\hbar^B(f)\in \gA_\hbar  \ (\hbar>0).
\end{align}
In other words, even though $f$ and $Q_\hbar^B(f)$ are completely different objects, for small $\hbar$ they are sufficiently close to each other and $\lim_{\hbar\to 0}Q_\hbar^B(f) = f$. This limit has to be interpreted in the sense that if one continuously follows the curve $\hbar\mapsto\sigma_f(\hbar)$ in the total space $\amalg_{\hbar\in I}A_\hbar$
of the bundle (equipped with the topology that makes this disjoint union a continuous bundle of $C^*$-algebras) to $\hbar=0$, one arrives at $f$.

We will see in  $\S \ref{applicationsschr}$ that this bundle plays an important role in the study of the classical limit of Schr\"{o}dinger operators, in the regime $\hbar\to0$.
}
\hfill$\blacksquare$
\end{example}

\begin{example}\label{spin1}
{\em
For any unital $C^*$-algebra $\gB$, we put
\begin{align}
&\gB_0^g=C(S(\gB)) \ \ (1/N=0); \label{B0} \\
&\gB_{1/N}=\gB^{\otimes N} \ \ (1/N>0), \label{BN}
\end{align}
where $\otimes^N$ denotes the $N$-fold projective tensor product of $\gB$ with itself (often called $\gB^N$ in what follows) and $S(\gB)$ is the algebraic state space of $\gB$ equipped with the weak $\mbox{}^*$-topology in which it is a compact convex set, e.g.\ the three-ball $S(M_2(\bC))\cong B^3\subset\bR^3$. Then, by \cite[Theorem 8.4]{Lan17} $\gB_0^{g}$ and $\gB_{1/N}$ may be turned into a continuous bundle of $C^*$-algebras $\gB^g$ over the base space $I=\{0\}\cup \{1/N \ |\  N\in \mathbb{N}\} \subset[0,1]$ (with relative topology, so that $1/N \to 0$ as $N\to\infty$ and where $\mathbb{N}=\{1,2,..,\}$). In order to define the continuous-cross sections we need the {\em symmetrization operator} $S_N : \gB^N \to \gB^N$, defined as the unique linear continuous extension of the following map on elementary tensors:
\begin{align}
S_N (a_1 \otimes \cdots \otimes a_N) = \frac{1}{N!} \sum_{\sigma \in {\cal P}(N)} a_{\sigma(1)} \otimes \cdots \otimes a_{\sigma(N)}. \label{defSN}
\end{align}
Furthermore, for $N\geq M$ we need to generalize the definition of $S_N$ to give a bounded  operator  $S_{M,N}: \gB^M 	\to \gB^N$, defined by linear and continuous extension of  
\begin{align}
S_{M,N}(b) = S_N(b \otimes \underbrace{1_{\gB} \otimes \cdots \otimes 1_{\gB} }_{N-M \mbox{\scriptsize times}}),\quad b \in \gB^M. \label{defSMN}
\end{align}
Given a sequence $(b)=(b_0,b_{1/N})_{N\in\bN}$ the part 
 $(b_{1/N})_{N\in\bN}$ away from zero (i.e.\ with $b_0$ omitted) is called {\bf symmetric} if there exist $M \in \mathbb{N}$ and $b_{1/M} \in \gB^{\otimes M}$ such that 
\begin{align}\label{one}
    b_{1/N} = S_{M,N}(b_{1/M})\:\mbox{for all }  N\geq M,
\end{align}
and {\bf quasi-symmetric} if  for every $N\in \mathbb{N}$ one has $b_{1/N} = S_{N}(b_{1/N})$,
and for every $\epsilon > 0$, there is a symmetric sequence $(c)$
as well as  $M \in\mathbb{N}$ (both depending on $\epsilon$) such that 
\begin{align} \| b_{1/N} - c_{1/N}\| < \epsilon\: \mbox{ for all } N > M.
\end{align} 
Now, if $(b)$ is a quasi-symmetric sequence, and $\omega$ is a state on $\gB$ then the following
limit exists \cite{RW}
\begin{align}\label{eq847}
b_0(\omega)=\lim_{N\to\infty}\omega^N(b_{1/N}),
\end{align}
where $\omega^N$ is the $N$-fold tensor product of $\omega$ with itself, defining a state on $\gB^N$. The ensuing function $b_0$ on the state space $S(\gB)$ is continuous, so that $b_0$ is an element of the algebra $\gB_0^{g}$.  The continuous cross-sections of $\gB^g$ correspond to quasi-symmetric sequences $(b)$ through
\begin{align}
&\sigma:0\mapsto b_0;\\
&\sigma:1/N\mapsto b_{1/N}  \ (1/N>0),
\end{align}
the former defined by \eqref{eq847}.
The algebra $\gB^{g}$ is also called the algebra of {\bf global} or {\bf quasi-symmeric} observables. This $C^*$-bundle plays an important role in the semi-classical behaviour of mean-field quantum spin systems defined on a lattice of $N$ sites, as $N\to\infty$. That is, we  put $\hbar=1/N$, where $N\in\mathbb{N}$ is interpreted as the number of sites of the model. In that case, one may take $\gB=M_k(\bC)$ for some $k\in\bN$ (see $\S \ref{applicationsCW}$).
}
\hfill$\blacksquare$
\end{example}

\begin{example}\label{spin2}
{\em
For any unital $C^*$-algebra $\gB$, we put
\begin{align}
 &\gB_0^{l}=\gB^{\infty} \ \ (1/N=0) \label{B0l};\\
 &\gB_{1/N}=\gB^{N} \ \ (1/N>0), \label{BN}
\end{align}
where $\gB^{\infty}$ is the infinite projective tensor product of $\gB$ with itself. The fibers $\gB_0^{l}$ and $\gB_{1/N}$ may be turned into a continuous bundle of $C^*$-algebras $\gB^{l}$ over the base space $I=\{0\}\cup 1/\mathbb{N}$ ($\mathbb{N}=\{1,2,..,\}$) \cite[Theorem 8.8]{Lan17}. In order to describe the cross-sections in this case, we have to realize the infinite tensor product $\gB_0^{l}$ as
equivalence classes of quasi-local sequences.  A sequence $(b_{1/N})_{N\in\bN}$ is called {\bf local} if there exist $M \in \mathbb{N}$ and $c_{1/M} \in \gB^{\otimes M}$ such that 
\begin{align}\label{one}
    b_{1/N} = c_{1/M}\otimes 1_\gB\cdots\otimes 1_\gB,
\end{align}
with $N-M$ copies of the unit $1_\gB\in \gB$.  A sequence $(b_{1/N})_{N\in\bN}$ is called {\bf quasi-local} of for every $\epsilon > 0$, there is a local sequence $(c_{1/N})_{N\in\mathbb{N}}$
and some $M \in\mathbb{N}$  such that 
\begin{align} \| b_{1/N} - c_{1/N}\| < \epsilon\: \mbox{ for all } N > M.
\end{align} 
Introduce an equivalence relation on the quasi-local sequences by saying that $(b)\sim (b')$ \  iff \ $\lim_{N\to\infty}||b_{1/N}-b'_{1/N}||=0$. The algebra $\gB^{\infty}$ consists of equivalence classes $[b]\equiv b_\infty$ of quasi-local sequences. These form a $C^*$-algebra under pointwise operations (in $N$) and norm $||b_\infty||=\lim_{N\to\infty}||b_{1/N}||$. Continuous cross-sections of $\gB^l$ then correspond to quasi-local sequences $(b)$ through
\begin{align}
&\sigma:0\mapsto b_\infty;\\
&\sigma:1/N\mapsto b_{1/N}  \ (1/N>0),
\end{align}
The algebra $\gB^{l}$ is called the algebra of {\bf quasi-local } observables. This $C^*$-bundle is typically used for studying the limit of large particle numbers of quantum spin systems with nearest neighbor interactions, like the quantum Ising or Heisenberg model of Example \ref{antiferroheisenberg}.
}
\hfill$\blacksquare$
\end{example}

\begin{example}\label{spin3}
{\em 
Let us consider the $(N+1)$-dimensional symmetric subspace of the Hilbert space $\bigotimes_{n=1}^N\mathbb{C}^2$,
\begin{align}
\text{Sym}^N(\mathbb{C}^2):= \underbrace{\mathbb{C}^2 \otimes_s \cdots  \otimes_s\mathbb{C}^2}_{N \: times}\subset\bigotimes_{n=1}^N\mathbb{C}^2,
\end{align}
where $\otimes_s$ indicates the symmetric tensor product.  Indicating the algebra of bounded operators on $\text{Sym}^N(\mathbb{C}^2)$ by $B(\text{Sym}^N(\mathbb{C}^2))$, it is known \cite[Theorem 8.1]{Lan17} that
 \begin{align}
 &\gA_0':=C(S^2); \label{B021}\\
 &\gA_{1/N}':= B(\text{Sym}^N(\mathbb{C}^2))\cong M_{N+1}(\bC),\label{BN21}
   \end{align}
are the fibers of a continuous bundle of $C^*$-algebras $\gA'$ over base space $I= \{1/N \:|\: N \in  \mathbb{N}\} \cup\{0 \}\equiv 1/\dot{\mathbb{N}}$, with $\dot{\mathbb{N}}=\mathbb{N}\cup\{\infty\}$ and $\mathbb{N}=\{1,2,..,\}$ as before.
Analogous to  Example \ref{schr} the continuous cross-sections are given by all sequences $(a_{1/N})_{N\in\dot{\mathbb{N}}}\in\Pi_{N\in\dot{\mathbb{N}}}\gA_{1/N}'$ for which $a_0\in C(S^2)$ and $a_{1/N}\in \gA_{1/N}'$  and  such that the sequence  $(a_{1/N})_{N\in\mathbb{N}}$ is asymptotically equivalent to $(Q_{1/N}'(a_0))_{N\in\mathbb{N}}$, in the sense that
\begin{align}
\lim_{N\to\infty}||a_{1/N}-Q_{1/N}'(a_0)||_N=0.\label{equivalencebookklaas}
\end{align}
Here, the symbol $Q_{1/N}'$ denotes the quantization maps
\begin{align}
Q_{1/N}':\gA_0' \to \gA_{1/N}',
\end{align}
which are  are defined by\footnote{Equivalent definitions of these quantization maps are used in literature, see e.g. \cite{Lan17,Pe72}. In particular, the quantization maps  \eqref{defquan3} define a Berezin quantization on $C(S^2)$ and are surjective onto $B(\text{Sym}^N(\mathbb{C}^2))$.} the integral computed in weak sense
\begin{align}\label{defquan3}
Q_{1/N}'(f)& :=
 \frac{N+1}{4\pi}\int_{S^2}f(\Omega)|\Psi_N^\Omega\rangle\langle\Psi_N^\Omega|d\Omega\:,
\end{align}
where $f$ denotes an arbitrary continuous function on $S^2$, $d\Omega$ indicates the unique $SO(3)$-invariant Haar measure on ${S}^2$ with $\int_{{S}^2} d\Omega = 4\pi$, and $|\Psi_N^\Omega\rangle\langle \Psi_N^\Omega|\in B(\text{Sym}^N(\mathbb{C}^2))$ are so-called $N$-coherent spin states defined below. To this end we use the bra-ket notation. 
Let $|\!\uparrow\rangle, |\!\downarrow\rangle$ be the eigenvectors of $\sigma_3$ in $\mathbb{C}^2$, so that
$\sigma_3|\!\uparrow\rangle=|\!\uparrow\rangle$ and $\sigma_3|\!\downarrow\rangle=- |\!\downarrow\rangle$, and where  $\Omega \in {S}^2$, with  polar angles  
$\theta_\Omega \in (0,\pi)$, $\phi_\Omega \in (-\pi, \pi)$, we then define the unit vector
\begin{align}\label{om1}
|\Omega\rangle_1  = \cos \frac{\theta_\Omega}{2} |\!\uparrow\rangle + e^{i\phi_\Omega}\sin   \frac{\theta_\Omega}{2} |\!\downarrow\rangle.
\end{align}
 If $N \in \mathbb{N}$, the associated {\bf $N$-coherent spin state} $\Psi_N^{\Omega}:=|\Omega\rangle_N\in \mathrm{Sym}^N(\mathbb{C}^2)$, equipped with the usual scalar product $\langle \cdot ,\cdot \rangle_N$ inherited from    
$(\mathbb{C}^2)^N$,  is defined as follows \cite{Pe72}:
\begin{align}\label{om2}
|\Omega\rangle_N =  \underbrace{|\Omega\rangle_1 \otimes \cdots \otimes |\Omega \rangle_1}_{N \: times}.
\end{align}
The $C^*$-bundle  $\gA'$ is suitable to study the semi-classical behaviour of the Bose-Hubbard model on a fixed finite lattice where the number of bosonic particles is sent to infinity, it can be used to analyze quantum spin systems in the high spin limit, i.e. the limit in the spin quantum number $J:=N/2\to\infty$ (see $\S$\ref{applicationsBH}), or it can be applied to permutation- invariant quantum spin systems limit or large particles \cite{Ven2020}.
}
\hfill$\blacksquare$
\end{example}
It is the bundle $\gB^l$ presented in the second example that connects to the frequently-called {\bf thermodynamic limit}, since the limiting algebra is higly non-commutative and usually plays a key role in studying local quantum statistical mechanics of infinite volume systems.\footnote{The mathematical approach for obtaining this limit exists under the name {\em inductive limit}.} Instead, the bundle algebra $\gB^g$, whose fibers for each finite $N$ are identical to those of $\gB^l$ but differ dramatically at $N=\infty$, i.e. for $1/N=0$, corresponding to the commutative $C^*$-algebra $\gB_0^{g}$, is the correct one to describe classical thermodynamics as limit of quantum statistical mechanics. Thus, it is precisely the choice of physical observables (quasi-local or quasi-symmetric ones) that determines the limiting theory in this case.  In turn, the limiting algebras $\gA_0^c$ and $\gA_0'$ corresponding to the bundles $\gA^c$ and $\gA'$, respectively, are commutative and therefore they relate to a classical theory as well. In other words, the bundle algebras $\gB^g$, $\gA^c$ and $\gA'$ are the appropriate ones to study the {\bf classical limit}, whose precise concept is outlined in $\S$\ref{sec:classicallimit}.

\subsection{Strict deformation quantization}\label{deformationquantization}

In this paper we emphasize the importance of a continuous bundle of $C^*$-algebras as an algebraic framework to study the emergent phenomena of spontaneous symmetry breaking. We shall particularly focus on $C^*$ bundles, each with a commutative $C^*$-algebra as fiber above $\hbar=0$, meaning that the limiting algebra describes a classical theory. In other words, we focus on the {\em classical} rather than the thermodynamic limit. A mathematically correct way to study this limit is by {\em (deformation) quantization}.\footnote{We shall mainly rely on the concept of (strict) deformation quantization developed in the 1970s  (Berezin \cite{Ber}; Bayen et al. \cite{Bay}), where non-commutative  algebras characteristic of quantum mechanics arise as deformations of commutative Poisson algebras characterizing classical theories. In Rieffel’s \cite{Rie89,Rie94} approach to deformation quantization, further developed by Landsman \cite{Lan98}, the deformed algebras are $C^*$-algebras, so that the apparatus of operator algebras becomes available.} 

The general idea of a deformation quantization is to consider a classical theory, whose observables are described by a class of sufficiently regular functions over a certain phase space $X$ (assumed to be locally compact and Hausdorff), as the zero-limit of a sequence of usually non-commutative or quantum theories labeled by a semi-classical parameter $\hbar$, whose observables are represented by self-adjoint operators on a corresponding sequence of Hilbert spaces. The non-commutative theories correspond to $\hbar>0$ whilst the classical theory, loosely speaking, is obtained by considering the limit $\hbar\to 0$ in a continuous manner. To be more precise this limit is established by means of a selection of sequences of observables (parametrized by $\hbar\geq 0$) with a suitable continuity property formulated in terms of $C^*$-algebras. Concretely, such sequences are nothing else than a subclass of the continuous cross-sections of a certain continuous bundle of $C^*$-algebras.\footnote{We have seen several examples in $\S$\ref{contbundleofc}.}

In this algebraic setting, as already indicated, the quantum observables are given by self-adjoint elements of abstract $C^*$-algebras $\gA_\hbar$ of formal operators $a\in \gA_\hbar$. The algebraic states in turn are complex-valued positive linear normalized functionals $\omega_\hbar : \gA_\hbar \to \bC$ with the physical meaning of $\omega_\hbar(a)$ as the expectation values of the observable $a=a^*$ in the state $\omega_\hbar$. A great advantage is that  this algebraic approach, differently from the usual Hilbert space formulation, is suitable even for classical theories. This is because  the set of (sufficiently regular) functions $f$ on the space of phases $X$ representing classical observables has a natural structure of a {\em commutative}  $C^*$-algebra $\gA_0=C_0(X)$, so that, due to the Riesz representation theorem algebraic states can be identified with probability measures over $X$, $\omega_0(f) = \int_X f d\mu_\omega$.\footnote{In general, the set $\gA_0$ can be any commutative $C^*$-algebra. This paper is however based on quantization of a Poisson manifold $X$, which naturally corresponds to the $C^*$-algebra $\gA_0=C_0(X)$, as indicated in \cite[Chapter 7]{Lan17}. In this setting,  strictly speaking, the classical observables should be elements of a dense $*$-Poisson subalgebra of $C_0(X)$ (which itself is not a Poisson algebra!), in order to define a Poisson bracket and therefore a classical theory.} 
Moreover we emphasize that, as a result of the famous GNS-construction, this more abstract viewpoint actually encompasses the Hilbert space formulation, so that the rather
abstract algebraic perspective can always be recast to a standard Hilbert space framework.

An instrument of absolute relevance in this framework is the notion of {\em quantization map}, which design can be traced back to Dirac. From a modern point of view it consists of a map $Q_\hbar : \gA_0 \ni f \mapsto Q_\hbar(f) \in \gA_\hbar$ which associates to classical observables $f\in \gA_0$ (or a substructure of it) quantum observables $Q_\hbar(f)\in \gA_\hbar$ and satisfies a number of various conditions. We now introduce the concept of a deformation quantization of classical structures, in particular of a Poisson manifold, adopted from \cite[Def. 7.1]{Lan17}. 

\begin{definition}\label{def:deformationq}
A {\bf deformation quantization}\footnote{Named  {\em continuous quantization}
of a Poisson manifold in  \cite[Definition II 1.2.5]{Lan98}.} of a Poisson manifold $(X, \{\cdot,\cdot\})$ consists of:
 \begin{itemize}
\item[(1)]  a {continuous  $C^*$-bundle} $(I, \gA, \pi_\hbar:\gA\to \gA_\hbar)$,
 where $I$ is a subset of $\mathbb{R}$ containing $0$ as accumulation point and $\gA_0=C_0(X)$ 
equipped with norm $||\cdot||_{\infty}$;
\item[(2)] a dense $*$-subalgebra $\tilde{\gA}_0$ of  $C_0(X)$ closed under the action of Poisson brackets 
(so that $(\tilde{\gA}_0, \{\cdot, \cdot\})$ is a complex Poisson algebra);
\item[(3)]  a  collection of   {\bf quantization maps} $\{Q_\hbar\}_{\hbar\in I}$, namely linear maps $Q_{\hbar}:\tilde{\gA}_0  \to \gA_{\hbar}$ 
(possibly defined on $\gA_0$ itself and next restricted to $\tilde{\gA}_0$)
such that: 
\begin{enumerate}
\item[(i)] $Q_0$ is the inclusion map $\tilde{\gA}_0 \hookrightarrow \gA_0$ (and $Q_{\hbar}(1\spa1_{\gA_0})=1\spa1_{\gA_{\hbar}}$
if  $\gA_0,$ and $\gA_\hbar$ are unital for all $\hbar \in I$);
\item[(ii)] $Q_{\hbar}(\overline{f}) = Q_{\hbar}(f)^*$, where $\overline{f}(x):=\overline{f(x)}$;
\item[(iii)] for each $f\in\tilde{\gA}_0$, the assignment
$
0\mapsto f, \quad 
\hbar\mapsto Q_{\hbar}(f)$ when $\hbar \in I\setminus \{0\},
$
defines  a continuous section of $(I, \gA, \pi_\hbar)$,
\item[(iv)]  each pair $f,g\in \tilde{\gA}_0$ satisfies the {\bf Dirac-Groenewold-Rieffel condition}:
\begin{align*}
\lim_{\hbar\to 0}\left|\left|\frac{i}{\hbar}[Q_{\hbar}(f),Q_{\hbar}(g)]-Q_{\hbar}(\{f,g\})\right|\right|_{\hbar}=0.
\end{align*}
\end{enumerate}
\end{itemize}
 If $Q_\hbar(\tilde{\gA}_0)$ is dense in $\gA_\hbar$ for every $\hbar \in I$, then the deformation quantization is called {\bf strict}.\\
(If $Q_\hbar$ is defined on the whole $C_0(X)$, all conditions except (iv) are assumed to be valid on $C_0(X)$.)
\end{definition}
Elements of $I$ are interpreted as possible values of Planck’s constant $\hbar$ and $\gA_{\hbar}$
 is the quantum algebra of observables of the theory at the given value of $\hbar\neq 0$. As a result of condition $(ii)$ in Definition \ref{def:deformationq}, for real-valued $f$ the operator
$Q_\hbar(f)$ is self-adjoint and therefore denotes the quantum observable associated to the classical observable $f$. 

It immediately follows from the definition of a continuous bundle of $C^*$-algebras that for any $f\in \tilde{\gA}_0$ the next continuity properties\footnote{In order to define a deformation  quantization it is not necessary to start from a continuous $C^*$-bundle, but it is sufficient to assign  
quantization maps satisfying some conditions \cite{Lan98}.} hold
\begin{itemize}
\item {\bf Rieffel's condition:}
\begin{align}\label{Rifc}
\lim_{\hbar\to 0}\|Q_{\hbar}(f)\|_\hbar=\|f\|_{\infty}\:;
\end{align}
\item {\bf von Neumann condition:}
\begin{align}\label{vNc}
\lim_{\hbar\to 0}\|Q_{\hbar}(f)Q_{\hbar}(g)-Q_{\hbar}(fg)\|_\hbar=0.
\end{align}
\end{itemize}
Indeed, the section $I \ni \hbar \mapsto Q_{\hbar}(f)Q_{\hbar}(g)-Q_{\hbar}(fg)$ is a continuous section because, it is constructed with the pointwise operations of the $C^*$-algebra $\gA$ and 
$(I, \gA, \pi_\hbar)$ is a {continuous  $C^*$-bundle}, finally $Q_{0}(f)Q_{0}(g)-Q_{0}(fg) = fg-fg=0$, hence (iii) in Definition \ref{def:continuous algebra bundle} implies (\ref{vNc}).
\\\\
We stress that the concept of deformation quantization can be studied in the context of tensor product bundles as well ~\cite{MV2}. This however goes beyond the scope of this paper. 
Let us now give some concrete examples some of them already introduced in $\S$\ref{contbundleofc}.

In Example \ref{schr}, the $C^*$-bundle $\gA^c$ over $I=[0,1]$ with Berezin maps $Q_{\hbar}^B$ defined by \eqref{defQB} gives a strict deformation quantization of $\bR^{2n}$, where $\bR^{2n}$ carries the canonical Poisson structure induced by the natural symplectic form $\sum_{k=1}^n dp_k \wedge dq^k$. The dense Poisson $*$-subalgebra of $C_0(\bR^{2n})$ can be chosen to  be all compactly supported smooth functions on $\bR^{2n}$, and $Q_{\hbar}^B$ restricted to this subalgebra surjectively maps onto a dense subalgebra of $B_\infty(L^2(\bR^n))$  \cite{MorVen2}.

Also Example \ref{spin1} relates to a deformation quantization, at least when $\gB=M_k(\bC)$ as proved in \cite{LMV}. Indeed, it can be shown that $S(M_k(\bC))$ is canonically a Poisson manifold with stratified boundary. In the special case that $k=2$, the Poisson bracket assumes the form
\begin{align}\label{pbra}
\{f,g\}^{(B^3)}({\bf x}) = \sum_{a,b,c =1}^{3}\epsilon_{abc}x_c \frac{\partial f}{\partial x_a}\frac{\partial g}{\partial x_b}\:, \quad {\bf x}\in B^{3},\:
\end{align}
where $S(M_2(\bC))\cong B^3=\{ \mathbf{x} \in \mathbb{R}^3 \:|\: \|\mathbf{x}\|\leq 1\}$, the closed unit ball in $\mathbb{R}^3$. This isomorhim is given by the well-known parametrization
\begin{align}\label{pardensitym}
\rho_{x,y,z}=\frac{1}{2}\begin{pmatrix} 1+z & x-iy\\ x+iy & 1-z \end{pmatrix},
\end{align}
of an arbitrary density matrix $\rho_{x,y,z}$ on $\bC^2$.
Quantization maps are defined by symmetric sequences through the maps \eqref{deformationqunatizaion} (see $\S\ref{applicationsCW}$). 
For further details we refer to \cite{LMV}.

Example \ref{spin3} yields a strict deformation quantization as well. Clearly the manifold $S^2$ is a symplectic (in particular, a Poisson) manifold with Poisson bracket induced by the symplectic form $\sin{\theta} d\theta\wedge d\phi$, where $\theta\in (0,\pi)$, and $\phi\in (0,2\pi)$. The Berezin quantization maps \eqref{defquan3} restricted to the dense Poisson $*$-subalgebra $C^\infty(S^2)$ of all smooth functions on $S^2$ satisfy all assumptions of Definition \ref{def:deformationq}, in particular they map surjectively onto a dense subalgebra of $B(\text{Sym}^N(\bC^2))$ \cite[Theorem 8.1]{Lan17}. 

\begin{remark}
{\em 
Note that the quantization maps of Examples \ref{schr} and \ref{spin3} are defined through coherent states. The quantization maps in Example \ref{spin1} are defined in a different way: no coherent states are involved in their definition. This relies on the fact that the manifolds $\bR^{2n}$ and $S^2$ are symplectic and admit an additional structure of a so-called {\em coherent pure state quantization} \cite{Lan98}.
}
\hfill$\blacksquare$
\end{remark}

\subsection{Classical limit}\label{sec:classicallimit}
The above ingredients allow us to introduce the concept of the classical limit.\footnote{We point out to the reader that these topics are partially related by approaches in semi-classical and microlocal analysis \cite{GriSjo,Hel,SJHE,MZ}. As opposed to the $C^*$-algebraic framework used in this paper,  such works typically rely on the properties of the underlying Hilbert space structure.} We hereto assume we are given a strict deformation quantization of a Poisson manifold $X$ according to Definition \ref{def:deformationq}. We denote by $\tilde{\gA}_0\subset \gA_0$ the commutative dense Poisson $*$-subalgebra of $\gA_0=C_0(X)$ corresponding to the fiber at $\hbar=0$, and $\gA_\hbar$ the quantum algebra above $\hbar>0$. Given  a sequence of quantization maps $Q_\hbar: \tilde{\gA}_0 \ni f \mapsto Q_\hbar(f) \in \gA_\hbar$, we say that a sequence of states $\omega_\hbar:\gA_\hbar\to\mathbb{C}$ is said to be have a {\bf classical limit} if the following limit exists and defines a state $\omega_0$ on $\tilde{\gA}_0$,
\begin{align}
\lim_{\hbar\to 0}\omega_\hbar(Q_\hbar(f))=\omega_0(f), \ (f\in \tilde{\gA}_0).
\end{align}
By construction, this approach provides a rigorous meaning of the convergence of algebraic quantum states $\omega_\hbar$  to classical states $\omega_0$ on the commutative algebra  on $\tilde{\gA}_0$, when $\hbar \to 0$. 

A special case of interest are the quantum algebraic (vector) states $\omega_\hbar(Q_\hbar):=\langle \psi_\hbar,  Q_\hbar(f)   \psi_\hbar \rangle$ induced by some normalized unit vector $\psi_\hbar$. The subscript $\hbar$ indicates that the unit vectors might depend on $\hbar$, which is for example the case when the $\psi_\hbar$ correspond to eigenvectors of a $\hbar$-dependent Schr\"{o}dinger operator $H_\hbar$, or in case of spin systems, to eigenvectors $\psi_N$ of quantum spin Hamiltonians $H_N$. These issues have been presented from a technical perspective in \cite{Ven2020,MorVen2}. In such works it has been shown that this $C^*$-algebraic approach offers a complete interpretation and rigorous notion of the classical limit of quantum systems, even though eigenvectors of such operators in general do not admit a limit in the pertinent Hilbert space.

We will see in Section \ref{Applications} that this notion of the classical limit furthermore allows to study spontaneous symmetry breaking as emergent phenomenon when passing from the quantum realm to the classical world by switching off the semi-classical parameter, e.g. $\hbar$ where $\hbar\to 0$, or $1/N$ where $N\to\infty$. 
\begin{remark}
{\em 
In view of the classical limit, the limit $N\to\infty$ (where $N$ denotes the number of lattice sites, spin particles, etc.) by definition now yields a classical theory encoded on a certain phase space $X$ with ensuing algebra of classical observables given by $C_0(X)$ (or, strictly speaking, a substructure of it). We stress that in literature the limit $N\to\infty$ is often referred to as thermodynamic limit, regardless of the nature of the limiting theory, i.e. commutative or non-commutative.
}
\hfill$\blacksquare$
\end{remark}

\section{Symmetry in algebraic quantum theory}\label{symminalgquant}
In this section we introduce the notion of spontaneous symmetry breaking in algebraic quantum theory. In particular, we see that SSB applies to commutative as well as non-commutative $C^*$-algebras, and therefore to classical and quantum theories. In the event that they are encoded by a continuous bundle of $C^*$-algebras (cf. Definition \ref{def:continuous algebra bundle}) this allows one to study SSB as a possibly emergent phenomenon in the classical limit. All that is needed are the continuity properties of the $C^*$-bundle specified by the continuous cross-sections. 

Let us now introduce the general $C^*$-algebraic context where the notion of spontaneous symmetry breaking takes place.
A $C^*$-{\bf dynamical system} $(\gA, \alpha)$ is a
 $C^*$-algebra $\gA$ equipped with a {\bf dynamical evolution}, that is, a one-parameter group of $C^*$-algebra automorphisms $\alpha:= \{\alpha_t\}_{t \in \bR}$ that is {\em strongly continuous}  in $\gA$:   the map $\bR \ni t \mapsto \alpha_t(a) \in \gA$ is continuous for every $a\in \gA$.

\subsection{Dynamical symmetry groups and ground states} \label{groundsymm}
If $(\gA, \alpha)$ is a  $C^*$-dynamical system and $\omega$ is an $\alpha$-invariant state, 
i.e., $\omega(a) = \omega(\alpha_t(a))$ for every $a\in \gA$ and $t\in \bR$,
there is a  unique one-parameter group of unitaries $U := \{U_t\}_{t\in \bR}$ which implements $\alpha$ in the GNS representation, i.e.,  $\pi_\omega(\alpha_t(a))= U^{-1}_t \pi_\omega(a) U_t$,  and leaves fixed the cyclic vector $U_t\Psi_\omega = \Psi_\omega$ (see e.g., \cite{Lan17, Mor}).  It follows that $U$ is strongly continuous in $\gB({\cal H}_\omega)$ as a consequence of the strong continuity of $\alpha$ in $\gA$ and the properties of the GNS construction. This allows us to give the definition of a ground state \cite{BR1,BR2,Lan17}. 
\begin{definition}\label{def:grounds}
A {\bf ground 
state} of a $C^*$-dynamical system $(\gA, \alpha)$ is an algebraic state  $\omega: \gA \to \bC$ such that 
\begin{itemize}
\item[(a)]
the state is $\alpha$-invariant, i.e, $\omega(\alpha_t(a)) = \omega(a)$ for all $t\in \bR$ and all $a\in \gA$,
 \item[(b)] the self-adjoint generator $H$ of the strongly-continuous 
 one-parameter  unitary group $U_t= e^{-itH}$
which implements $\alpha$ in a given  GNS representation $({\cal H}_\omega, \pi_\omega, \Psi_\omega)$ under the requirement $U_t\Psi_\omega = \Psi_\omega$, has spectrum $\sigma(H) \subset [0,+\infty)$. 
\end{itemize}
\end{definition}
\noindent
The set  $S^{ground}(\gA, \alpha)$ of ground states of $(\gA, \alpha)$ is convex and $*$-weak closed,  so that it is also compact for the Banach-Alaoglu theorem. As a result of the Krein-Milman theorem, all ground states can be constructed out of limit points of convex combinations of  {\em extremal ground states}  in the $*$-weak topology. Even though the extremal ground states are the building blocks for constructing all other ground states, they are not necessarily {\em pure states}, i.e.,  extremal  states in the convex $*$-weak  compact set of {\em all} algebraic states on $\gA$. Nonetheless,  in many cases of physical interest extremal ground states exactly correspond to pure states  which are also ground states.

\subsection{Commutative case}

Definition \ref{def:grounds} applies in particular to the commutative case where $\gA := C_0(X)$ endowed with the $C^*$-norm $||\cdot||_\infty$, referred to a locally compact Hausdorff space $X$ possibly endowed with a Poisson structure $(C^\infty(X), \{\cdot, \cdot\})$. 
In this case the states $\omega$ are nothing but the regular\footnote{All positive Borel measures on $X$ are automatically regular due to Theorem 2.18  in \cite{Rudin}.} Borel probability measures $\mu_\omega$ over $X$.  More precisely, if $\omega : \gA \to \bC$ is an algebraic  state, the $C_0(X)$ version of  Riesz's representation theorem of generally complex  measures on locally compact Hausdorff spaces \cite{Rudin}, proves that $({\cal H}_\omega, \pi_\omega, \Psi_\omega)$ assumes the form
$${\cal H}_\omega = L^2(X, \mu_\omega)\:, \quad  (\pi_\omega(f)\psi)(\sigma) =f(\sigma) \psi(\sigma)\:,\quad 
 \Psi_\omega(\sigma) = 1\:,  \quad \mbox{for all $f\in C_0(X)$, $\psi\in 
{\cal H}_\omega$ and  $\sigma \in X$.}$$
 With this representation, the pure states are {\em Dirac measures} concentrated at any point $\sigma \in X$.

A $C^*$-dynamical system structure is constructed when  the dynamical evolution is furnished by the pullback action of the Hamiltonian flow  $\phi^{(h)}$,
 provided it is {\em  complete}, generated by a (real) hamiltonian function $h\in C^\infty(X)$, i.e., 
$\alpha^{(h)}_t(f) := f \circ \phi_t^{(h)}$ for every $f\in C_0(X)$ and $t\in \bR$. 

It is easy to prove that $(C_0(X), \alpha^{(h)})$ is a $C^*$-dynamical system (in particular $\alpha^{(h)}$ leaves $C_0(X)$ invariant and is strongly continuous \cite{MorVen2}).  
We have the following result on ground states \cite[Prop. 6.3]{MorVen2}.

\begin{proposition}\label{propNh3}
The ground states of $(C_0(X), \alpha^{(h)})$ are all of the regular Borel  probability measures on $X$ whose support is contained in the closed set $N_h := \{\sigma \in X \:|\: dh(\sigma) =0\}$.  
\end{proposition}

If $\omega$ is a ground state of $(C_0(X), \alpha^{(h)})$, in view of the above discussion,  $\alpha^{(h)}$ is trivially implemented:  $U_t = I$ for every $t\in \bR$ and the positivity 
 condition on  the spectrum of the generator of $U_t$ is automatically fulfilled. In this case the extremal elements of $S^{ground}(\gA, \alpha)$ are 
the Dirac measures concentrated at the points $\sigma\in X$ such that $dh(\sigma)=0$. In particular {\em extremal} ground states 
are {\em pure} states.  

\subsection{Weak and spontaneous symmetry breaking}\label{weakandssb}

When $(\gA, \alpha)$ is a $C^*$-dynamical system also endowed with a group $G$ acting on $\gA$ with a group representation $\gamma : G \ni g
 \to \gamma_g$ in terms of $C^*$-automorphisms $\gamma_g : \gA \to \gA$, we say that $G$ is a {\bf dynamical symmetry group} if
$\gamma_g \circ \alpha_t = \alpha_t \circ \gamma_g\quad \mbox{for all $g\in G$ and $t\in \bR$.}$ This leads to the following definition \cite[Def. 10.3]{Lan17}.

\begin{definition}\label{def:SSB}
Let $(\gA,\alpha)$ be $C^*$ dynamical system, and suppose have a (topological) group $G$ and a (continuous) homomorphism $\gamma: \to \text{Aut}(\gA)$ which is a symmetry of the dynamics. We say that the $G$-symmetry is {\bf spontaneously broken} (at $T=0$) if 
\begin{align}
(\partial_eS^{ground}(\gA))^G=\emptyset.
\end{align}
As before,  $S^{ground}(\gA)$ denotes the convex set of algebraic ground states of $\gA$, and $\partial_eS^{ground}(\gA)$ is the set of extremal ground states ($\partial_e$ indicates the extreme boundary). Finally, $\mathscr{S}^G:=\{\omega\in \mathscr{S}\ | \ \omega\circ\gamma_g=\omega \  \forall g\in G\}$ defined for any subset $\mathscr{S}\subset S(\gA)$ is the set of $G$-invariant states in $\mathscr{S}$.
\end{definition}
Therefore, spontaneous symmetry breaking  occurs for a dynamical system $(\gA, \alpha)$ endowed with  a dynamical symmetry group $G$ if there are no $G$-invariant  ground states which are extreme points in $S^{ground}(\gA, \alpha)$. Within the usual situation where the extremal points in  $S^{ground}(\gA, \alpha)$ are the ground pure states of $\gA$, occurrence of SSB means that $G$-invariant ground states {\em must be necessarily mixed states}. A more frequent situation happens if $(\partial_eS^{ground}(\gA))^G\neq \partial_eS^{ground}(\gA, \alpha)$, there is at least one extreme point in $S^{ground}(\gA, \alpha)$ that is not $G$-invariant. In this case one says that {\em weak symmetry breaking} takes place.

Furthermore, Definition \ref{def:SSB} extends to primary KMS states at inverse temperature $\beta\in(0,\infty)$. These states play an important role in the definition of pure thermodynamic phases. 

\begin{remark}\label{weakversusspon}
{\em 
The very definition of a ground state in the commutative case shows that, as opposed to what is physically accepted, not only absolute minima might be ground states, but all extremal points $\sigma$ on which $dh$ vanishes. In view of Definition \ref{def:SSB} spontaneous symmetry breaking would almost never occur: if the classical Hamiltonian $h$ has a maximum (as happens in physically relevant models) then always weak symmetry breaking, rather than spontaneous symmetry breaking takes place. In order to circumvent this problem we proceed as in \cite{Lan17,LMV} and only focus on absolute minima, i.e. we consider classical ground states as being points in phase space on which the classical Hamiltonian assumes an absolute minimum.
}
\hfill$\blacksquare$
\end{remark}

\section{Applications}\label{Applications}
In this section we discuss various physical models including their classical limits for the particular choice of ground states. As already mentioned, the quantization map will play a crucial role in order to connect quantum operators with their classical counterparts, and it particularly defines the set of physical observables of the quantum system. \footnote{Clearly, in order to obtain physically relevant results this set should not be too ``small''. In the examples presented in this paper this is obviously not the case.}
Moreover we study several cases in which a symmetry group is present, and emphasize the role of spontaneous symmetry breaking. The main idea is that if an algebraic quantum state is invariant under a certain $G$-action, then the limiting state is $G$-invariant as well, which at least at the level of ground states studied in our examples, yields spontaneous symmetry breaking in the classical limit. All these ideas and concepts are carefully explained and illustrated with several examples presented in the next  paragraphs.

\subsection{Schr\"{o}dinger operators}\label{applicationsschr}

Consider a sequence of $\hbar$-dependent Schr\"{o}dinger operators $H_{\hbar}$ defined on some dense domain of $\mathcal{H}=L^2(\mathbb{R}^n, dx)$. Such operators are typically given by
\begin{align}
H_{\hbar}:=\overline{-\hbar^2\Delta + V}, \quad \hbar>0\:,\label{Schroper}
\end{align}
where $\Delta$ denotes the Laplacian on $\mathbb{R}^{n}$, and $V$ denotes multiplication by some real-valued function on $\mathbb{R}^n$, playing the role of the potential. 
Our general hypotheses on $V$ will be the following ones.
\begin{itemize}\label{HypotV}
\item[{\bf (V1)}] $V$ is a  real-valued $C^\infty(\bR^n)$ function.
\item[{\bf (V2)}] $ V(x) \to +\infty $ for $|x| \to +\infty$ (i.e., for every $M>0$, there is $R_M>0$
 such that $V(x) > M$ if $|x|>R_M$).
\item[{\bf (V3)}] $e^{-tV} \in {\cal S}(\bR^n)$ for $t>0$ (the set ${\cal S}(\bR^n)$ is the Schwartz space on $\bR^n$).
\end{itemize}
As a result of these hypotheses, standard arguments show that the operator  $-\hbar^2\Delta + V$ is essentially-self adjoint  on $C_0^\infty(\bR^n)$  and that $H_\hbar\geq 0$  (see e.g. \cite[Theorem X.28]{RS2}). In particular, the resolvent of $H_\hbar$ is compact  \cite[Theorem XIII.67]{RS4}. According to standard results on positive compact operators (see e.g \cite{Mor}), if $\hbar>0$,
\begin{itemize}
\item[(a)]  the spectrum of $H_\hbar$ is a pure point spectrum and there is a corresponding Hilbert basis of eigenvectors $\{\psi^{(j)}_\hbar\}_{j=0,1,\ldots}$
with corresponding eigenvalues
\beq \sigma(H_\hbar) = \{E^{(j)}_\hbar\}_{j=0,1,2,\ldots}\quad \mbox{with $0 \leq E^{(j)}_\hbar \leq E^{(j+1)}_\hbar \to +\infty$ as $j\to +\infty$,}\label{specH}\eeq
where  every eigenspace has finite dimension;
\item[(b)]  $e^{-tH_\hbar}$ 
is compact with  spectrum 
$\sigma(e^{-tH_\hbar}) = \{0\}\cup  \{e^{-tE^{(j)}_\hbar}\}_{j=0,1,2, \ldots}\:,$
 $0$ being the unique point of the continuous spectrum, and the  eigenspaces of $H_\hbar$ and $e^{-tH_\hbar}$ coincide;  
\item[(c)]  the minimal eigenvalue $E^{(0)}_\hbar$ of $H_\hbar$ corresponds to the maximal eigenvalue of  $e^{-tH_\hbar}$
according to
\begin{align}
e^{-t E_\hbar^{(0)}} = ||e^{-tH_\hbar} ||.
\end{align}
\end{itemize}
Furthermore, as a result of \cite[Prop. 4.2]{MorVen2}, the following properties hold
\begin{itemize}
\item[(d)] $||e^{-t H_\hbar}|| \to e^{-t \min V}$ as $\hbar \to 0^+$.
\item[(e)]  The eigenspace of  $H_\hbar$ associated to the minimal  eigenvalue  $E^{(0)}_\hbar$ has dimension $1$ for any $\hbar>0$.
\end{itemize}
This leads to a theorem proving the classical limit for the ground state of such Schr\"{o}dinger operators as $\hbar\to 0$ \cite[Theorem 5.4]{MorVen2}.\footnote{We stress that under a certain scale separation the physical meaning of the limit in Planck's constant $\hbar\to 0$ can be interpreted as the limit $m\to\infty$, where $m$ denotes the mass of the quantum particle. Indeed, the limit $\hbar\to 0$ where $\hbar$ occurs as $-\frac{\hbar^2}{2m}$ in front of the Laplacian at fixed mass (usually set to $1$) of a general Schr\"{o}dinger operator can be equivalently obtained by sending $m$ to infinity at fixed $\hbar$.} 

\begin{theorem}\label{thm:claslimNEWSchr}
Consider a group $G$ either finite or topological compact,  a selfadjoint Schr\"odinger operator on $L^2(\bR^n, dx)$
 $H_\hbar:=\overline{ -\hbar^2 \Delta + V},$
as in (\ref{Schroper}) where  $V: \bR^n \to \bR$ satisfies  (V1)-(V3),
and  assume the following hypotheses.
\begin{itemize} 
\item $G$ acts, continuously in the topological-group case\footnote{The action $G\times \mathbb{R}^{2n} \ni
 (g,\sigma) \mapsto g\sigma \in \mathbb{R}^{2n}$ is continuous.}, 
on  $(\mathbb{R}^{2n}, \sum_{k=1}^n dp_k \wedge dq^k)$ in terms of symplectomorphisms. 
\item Defining $h(q,p) :=p^2 + V(q)$, the action of $G$ leaves invariant $V$\footnote{Since the minimum set $\Sigma_{\min h}$ in particular corresponds to $p=0$, one would perhaps expect to require that $G$ leaves invariant only the minimum set of $V$. However, in this case it might happen that $H_\hbar$ does not commute with the unitary representation $U_g$ defined by  $U_g\phi(x)=\phi(g^{-1}x)$. As a result, $G$ may not define a dynamical symmetry group, which is necessary for the discussion of SSB. This problem is avoided be requiring that $V$ is $G$-invariant.} and it acts transitively on $\Sigma_{\min h}=h^{-1}(\{\min h\})$.
\end{itemize}
  Then the following facts are valid  for every chosen $\sigma_0\in \Sigma_{\min h}$ and for 
a family $\{\psi^{(0)}_\hbar\}_{\hbar>0}$ of (normalized) eigenvectors of
$H_\hbar$
   with  minimal (and thus automatically non-degenerate) eigenvalues $\{E^{(0)}_\hbar\}_{\hbar>0}$ converging to $\min_{q\in \bR^n} V(q)= \min_{(q,p)\in \bR^{2n}} h(q,p)$ as $\hbar\to 0$.\footnote{This family always exists as a result of property $(d)$ above.}
\begin{itemize}
\item[(1)] If $G$ is topological and compact,
\begin{align}
\lim_{\hbar\to 0^+}\langle\psi^{(0)}_\hbar,Q_{\hbar}^B(f)\psi^{(0)}_\hbar\rangle=\int_{G} f(g\sigma_0)d\mu_G(g),  \quad  \mbox{for every $f\in C_0(\mathbb{R}^{2n})$;}\label{classical limit22}
\end{align}
where $\mu_{G}$ is the normalized Haar measure of $G$.
\item[(2)] If $G$ is finite,
\begin{align}
\lim_{\hbar\to 0^+}\langle\psi^{(0)}_\hbar,Q_{\hbar}^B(f)\psi^{(0)}_\hbar\rangle=\frac{1}{N_G} \sum_{g\in G} f(g \sigma_0),  \quad  \mbox{for every $f\in C_0(\mathbb{R}^{2n})$;}\label{classical limit2b2}
\end{align}
where $N_G$ is the number of elements of $G$. The operator $Q_\hbar^B(f)$ denotes the Berezin quantization maps introduced in Example \ref{schr}.
\end{itemize}
The left and right-hand sides of (\ref{classical limit22}) and (\ref{classical limit2b2}) are independent of the choice of $\sigma_0$.
\end{theorem}

\begin{remark}
{\em
The proof of the theorem is based on the algebraic properties of the Berezin quantization maps. Indeed, the crucial idea of the proof is to show that $Q_{\hbar}^B(e^{-h})$ is in some sense a  ``good'' approximation of $e^{-H_\hbar}$, where $h(q,p)=p^2+V(q)$. The semiclassical behaviour of the Schr\"{o}dinger operator is therefore transferred to Berezin quantization of the function $e^{-h}$, which is in turn well-manageable in the limit $\hbar\to 0$ using the continuity properties of the maps $Q_\hbar^B$.  We refer to \cite{MorVen2} for further reading.
}
\hfill$\blacksquare$
\end{remark}

This theorem allows us to discuss the role of symmetry breaking in a complete algebraic manner. To simplify the discussion we focus on the case $V(q)=(q^2-1)^2 \ \ (q\in \bR^n)$. Clearly, this potential satisfies (V1)-(V3), and moreover, it is not difficult to see that all the hypotheses of the Theorem \ref{thm:claslimNEWSchr} are satisfied for $G=\bZ_2$ if $n=1$, or $G=SO(n)$ if $n>1$, which from now on, will be the groups of consideration. 
\\\\
To discuss the role of SSB, we start on the classical site, i.e. $\hbar=0$. We hereto take the $C^*$-algebra $C_0(\bR^{2n})$ and consider the time evolution $\alpha^{(h)}$ generated by the Hamiltonian flow $\phi^{(\hbar)}$ induced by the Hamiltonian $h = p^2 + V(q)$.  It is not difficult to see that $G$ is a dynamical symmetry group of $(C_0(\bR^{2n}), \alpha^{(h)})$ with action 
\begin{align}
\gamma_g f := f \circ g^{-1}; \quad \mbox{for all $g\in G$ and $f\in C_0(\bR^{2n})$},\label{gammaf}
\end{align}
where $G=SO(n)$ (or $\bZ_2$) acts in the obvious way on $\bR^{2n}$, i.e. on each of the two factors $\bR^n$  separately, by rotation (as $n>1$) and  by reflexion (in case of $\bR^2$ corresponding to $G=\bZ_2$).
Indeed, the Hamiltonian flow $\phi^{(\hbar)}$ is complete since the level sets of $h$ are compact and every solution of Hamilton equations is contained in one such set as $h$ is dynamically conserved. Since the action of $G$ is given by symplectomorphisms and every $\gamma_g$ leaves $h$ invariant and thus it commutes
with the Hamiltonian flow, the result follows.
\\\\
Moreover, the classical ground states exhibit spontaneous symmetry breaking. To see this, one observes that the extremal ground states are defined by the Dirac measures concentrated at the set of zeros of $dh$.  In all cases the only $G$-invariant extremal ground state is located at $(q_0,p_0) = (0,0)$. In view of Remark \ref{weakversusspon} we can forget about this point since it is a maximum. There is however an infinite number of non $G$-invariant extremal ground states located at the points $(q, 0)$ with $|q|=1$ if $n>1$ and exactly two non $SO(n)$-invariant extremal ground states $(\pm1, 0)$ if $n=1$. For $|q|=1$ and $p=0$ it follows that the associated functionals  $\omega_{|q|=1,p=0}^{(0)}$ on $C_0(\bR^{2n})$, given by $\omega^{(0)}_{|q|=1,p=0}(f)=f((q,0))$ are not invariant under the symmetry \eqref{gammaf}, and thus the $SO(n)$-symmetry is spontaneously broken. A similar result holds of course for the functionals associated to the points $(\pm1, 0)$ yielding spontaneous symmetry breaking of $\bZ_2$.
\\\\
On the quantum side (i.e. for $\hbar>0$ with observable algebra $B_\infty(L^2(\bR^n))$), the dynamical  evolution described by a $\hbar$-parametrized family of  one-parameter group of $C^*$-automorphisms  $\bR \ni t \mapsto \alpha^{\hbar}_t$ is provided   by  a corresponding $\hbar$-parametrized family  of one-parameter unitary
groups $\bR \ni t \mapsto U^\hbar_t := e^{-it H_\hbar}$:
\begin{align}
\alpha^\hbar_t(A) := U^\hbar_{-t} A U^\hbar_t, \quad \mbox (A \in B_\infty(L^2(\mathbb{R}^n))\:). \label{alphahbar}
\end{align}
 It can be shown that $\alpha^\hbar$ is strongly continuous in $B_\infty(L^2(\mathbb{R}^n))$ \cite[Prop 6.2]{MorVen2}, and thus $(B_\infty(L^2(\mathbb{R}^n)), \alpha^\hbar)$ is a $C^*$-dynamical system.

Since the resolvent of $H_\hbar$ is compact, in particular, $H_\hbar$ is an observable of the physical system represented by $\gA_\hbar$. That is not the whole story because, it turns out that $G = \bZ_2$, for $n=1$, or $G= SO(n)$, if $n>1$,  becomes a 
dynamical symmetry group. Indeed, for each $g\in G$ the unitary operator $U_g:L^2(\bR^n)\to L^2(\bR^n)$ given by $U_g\phi(x)=\phi(g^{-1}x)$ induces an automorphism $\gamma_g\in \text{Aut}(B_\infty(L^2(\mathbb{R}^n)))$ given by
\begin{align}\label{Gaction}
\gamma_g(A)=U_gAU_g^*, \ \ (A\in B_\infty(L^2(\mathbb{R}^n))).
\end{align}
It is not difficult to see that, for each $\hbar>0$ the $G$-invariance of the model is expressed by the property
\begin{align}
\alpha_t^\hbar\circ\gamma_g=\gamma_g\circ \alpha_t^\hbar,
\end{align}
and thus this gives the symmetry of the dynamics.\footnote{This follows from the following fact. Since the potential $V$ is assumed to be $G$-invariant, the unitary operator $U_g$ commutes with $H_\hbar$,  and therefore also with $U_t^{\hbar}$.} Differently from the classical ($\hbar=0$) case here no SSB occurs. Indeed, the ground state eigenvector of the quantum Hamiltonian $H_\hbar$  is a unit vector $\Psi_\hbar^{(0)}$ on $L^2(\bR^n)$ for which the corresponding eigenvalue $E_\hbar^{(0)}$ lies at the bottom of the spectrum $\sigma(H_\hbar)$. Algebraically, such a unit vector $\Psi_\hbar^{(0)}$ defines a state $\omega_\hbar^{(0)}$ on the $C^*$-algebra of observables $\gA_\hbar= B_\infty(L^2(\bR^n))$, viz.
\begin{align}\label{alground}
\omega_\hbar^{(0)}(A)=\langle\Psi_\hbar^{(0)},A\Psi_\hbar^{(0)}\rangle  \ \ (A\in B_\infty(L^2(\bR^n))), 
\end{align}
where $\langle \cdot,\cdot\rangle$ is the inner product on $L^2(\bR^n)$. In the case of compact operators it can be shown that the definition of the ground state (Definition \ref{def:grounds} of $\S\ref{groundsymm}$) is equivalent to \eqref{alground} \cite[Prop. 6.5]{MorVen2}, \cite{vandeVen}.
Now, since $U_g$ commutes with $H_\hbar$ and the eigenvector $\Psi_\hbar^{(0)}$ corresponds to a non-degenerate minimum, it follows that
\begin{align}
\omega_{\hbar}^{(0)}\circ\gamma_g=\omega_{\hbar}^{(0)} \ \ (g\in G).
\end{align}
Therefore, the pure state $\omega_{\hbar}^{(0)}$ is $G$-invariant. This should be physically evident since there is only one ``ground state'' (in the sense of a vector state with minimal energy) which is $G$-invariant. 
\\\\
The (unique) algebraic ground state  $\omega_\hbar^{(0)}(\cdot)=\langle\Psi_\hbar^{(0)},(\cdot)\Psi_\hbar^{(0)}\rangle$ is $G$-invariant and pure, and converges (by Theorem \ref{thm:claslimNEWSchr}) to the $G$- invariant, but {\em mixed} state $\omega_0^{(0)}(f):=\int_Gf(g\sigma_0)d\mu_G(g)$ (or similarly, in case of $G=\bZ_2$, to the state defined by \eqref{classical limit2b2}) which also qualifies a ground state of $(C_0(\bR^{2n}),\alpha^{(h)})$ \cite{MorVen2}. According to Definition \ref{def:SSB} and the discussion above the state $\omega_0^{(0)}$ breaks the $G$-symmetry. We conclude that this algebraic framework shows that spontaneous symmetry breaking occurs as {\em emergent} phenomenon when passing from the quantum realm (no SSB) to the classical world (existence of SSB) by switching off $\hbar$.

\subsection{Mean-field theories}\label{applicationsCW}

In this section we consider the quantum Curie-Weiss model\footnote{This model exists in both a classical and a quantum version and is a mean-field approximation to the Ising model. See e.g. \cite{FV2017} for a mathematically rigorous treatment of the classical version, and \cite{CCIL} for the quantum version. For our approach the papers \cite{Bona,DW,RW}  
played an important role. See also \cite{ABN} for a very detailed discussion of the quantum Curie--Weiss model.}, which
is an exemplary quantum mean-field spin model. We stress that similar results hold for general mean-field quantum spin systems, whose details have been proved in \cite{Ven2020}. The {\bf quantum Curie Weiss} defined on a lattice with $N$ sites\footnote{The geometric configuration including its dimension is irrelevant, as is typical for mean-field models \cite{VGRL18}, so that we may as well consider the model in one dimension, i.e. defined on a chain.} is
  \begin{align}
H^{CW}_{1/N}: &  \underbrace{\mathbb{C}^2 \otimes \cdots  \otimes\mathbb{C}^2}_{N \: times}  \to 
\underbrace{\mathbb{C}^2 \otimes \cdots  \otimes\mathbb{C}^2}_{N \: times}; \\
H^{CW}_{1/N} &=-\frac{J}{2N} \sum_{i,j=1}^N \sigma_3(i)\sigma_3(j) -B \sum_{j=1}^N \sigma_1(j). \label{CWham}
\end{align}
Here $\sigma_k(j)$ stands for $I_2 \otimes \cdots \otimes \sigma_k\otimes \cdots \otimes I_2$, where $\sigma_k$ denotes the spin-Pauli matrix $\sigma_k \ (k=1,2,3)$ and occupies the $j$-th slot, and 
 $J,B \in \mathbb{R}$ are given constants defining the strength of the spin-spin coupling and the (transverse) external magnetic field, respectively. In terms of the macroscopic average spin operators
\begin{align}
T_\mu=\frac{1}{2N}\sum_{i=1}^N\sigma_\mu(i), \ \ (\mu=1,2,3),
\end{align}
the Hamiltonian \eqref{CWham} assumes the more transparent form
\begin{align}
H^{CW}_{1/N}=-2N(JT_3^2+BT_1).
\end{align}
Perhaps surprisingly, the quantum Curie-Weiss Hamiltonian has a classical counterpart on the Poisson manifold $S(M_2(\bC))$.  
To see where this classical counterpart comes from, we rely on the fact \cite{LMV} that the manifold  $S(M_2(\bC))$ admits a deformation quantization, with quantization maps $Q_{1/N}:\tilde{\gA}_0^g\subset C(S(M_2(\bC)))\to M_2(\bC)^{\otimes N}$ given by symmetric sequences explained in $\S\ref{spin1}$, and where  $\tilde{\gA}_0^g$ is a dense Poisson $*$-subalgebra of $C(S(M_2(\bC))$)  . To define these maps explicitely, we first identify $S(M_2(\bC)) \cong B^3$ (cf. \eqref{pardensitym}). Quantization maps are then defined on the dense $*$-subalgebra of $\tilde{\gA}_0^g\subset C(B^3)$ given by all polynomials  $p(x,y,z)$ in three real variables restricted to $B^3$, which can be shown to admit a Poisson structure \cite{LMV}. Consequently, we rely on the fact that each homogeneous polynomial  $p$ of degree $N$ uniquely corresponds to symmetrized tensor product of the form \cite[$\S$ 3.1]{LMV}
\begin{align}
 b_{j_1} \otimes_s\cdot\cdot\cdot\otimes_sb_{j_N}, 
\end{align}
where $b_{j_i}=\sigma_{i}$, ($i=1,..,3$). Under this identification, i.e. $p\leftrightarrow b_{j_1} \otimes_s\cdot\cdot\cdot\otimes_sb_{j_N}$, quantization maps $Q_{1/N}$ are defined by linear extension of its values on such $p$ in the following way \cite[$\S$ 3.2]{LMV}:
\begin{align}\label{deformationqunatizaion}
 &Q_{1/N}(p) =
\begin{cases}
    S_{L,N}(b_{j_1}\otimes_s\cdot\cdot\cdot\otimes_sb_{j_L}), &\ \text{if} \ N\geq L; \\
    0, & \ \text{if} \ N < L,
\end{cases}\\
&Q_{1/N}(1_{B^3}) = \underbrace{I_2 \otimes \cdots \otimes I_2}_{\scriptsize N \: times}. \label{deformationqunatizaion2}
\end{align}
Here, $S_{L,N}$ denotes the symmetrizer defined by \eqref{defSMN}.
\\\\
By definition,
 \begin{equation}
 H^{CW}_{1/N} \in \text{Sym}(M_2(\mathbb{C})^{\otimes N}),
\end{equation}
where $\text{Sym}(M_2(\mathbb{C})^{\otimes N})$ is the range of the symmetrizer $S_N$. In order to find the classical counterpart, we first normalize\footnote{This idea goes back to Lieb \cite{Lieb} (see also \cite{Lan17,LMV,Ven2020,VGRL18}).} $H_{1/N}^{CW}$ as  $H_{1/N}^{CW}/N$ and, based on a combinatorial argument, we may rewrite
\begin{align}
H^{CW}_{1/N}/N&= -\frac{J}{2N(N-1)} \sum^N_{i \neq j, \:i,j=1} \sigma_3(i)\sigma_3(j) - \frac{B}{N} \sum_{j=1}^N \sigma_1(j) + O(1/N).\nonumber \\
&= Q_{1/N}(h^{CW}_0)  + O(1/N),\label{QNh}
\end{align}
where $O(1/N)$ is meant in norm (i.e. the operator norm on each space $M_2(\mathbb{C})^{\otimes N}$), and
the {\bf  classical Curie--Weiss Hamiltonian} is 
\begin{align}
h^{CW}_0: B^3 &\mapsto\mathbb{R};\\
h^{CW}_0(x,y,z)  &= -\left(\frac{J}{2}z^2 + Bx\right), \quad \mathbf{x}=(x,y,z) \in B^3. \label{hcwc}
\end{align}
Therefore, up to a small error as $N\to\infty$,  the quantum  Curie--Weiss  Hamiltonian \eqref{CWham} is given by quantization of its classical counterpart \eqref{hcwc}. 
\\\\
To find the classical limit, we a priori have to fix $B\in (0,1)$ in which regime the CW model exhibits spontaneous symmetry breaking (SSB) (see the discussion below and \cite{Lan17,LMV,VGRL18}). For convenience, we set $J=1$. For this choice of parameters we consider the absolute minima of the classical CW hamiltonian. By a simple calculation these minima are attained in
\begin{align}
\mathbf{x_{\pm}}=(B,0,\pm\sqrt{1-B^2}), \label{sphericalminima}
\end{align}
and lie on the extreme boundary $S^2=\partial_eB^3$, the unit two-sphere in $\bR^3$. 

The above observations yield the the following theorem \cite[Thm. 3.4]{LMV} proving the existence of the classical limit in the number of sites $N$ of a sequence of ground state eigenvectors associated to the quantum Curie-Weiss model.

\begin{theorem}\label{mainsecond2}
Let  $Q_{1/N}: \tilde{\gA}_0^g \to M_2(\bC)^N$  be the quantization maps defined by linear extension of \eqref{deformationqunatizaion} - \eqref{deformationqunatizaion2}, and let $\Psi_N^{(0)}$ be the (unit) ground state eigenvector of the  Hamiltonian \eqref{CWham} of the quantum Curie--Weiss model. 
Then
\begin{equation}
\lim_{N\to\infty}\omega_{1/N}^{(0)}( Q_{1/N}(f))=\frac{1}{2}(f(\mathbf{x}_+)+f(\mathbf{x}_-)),\label{question}
\end{equation}
for any polynomial function $f$ on $B^3$ (parametrizing the state space of $M_2(\bC)$),
where the points $\mathbf{x}_\pm\in B^3$ are given by \eqref{sphericalminima}.
\end{theorem}

Finally, we explain how symmetry breaking plays a role in the Curie-Weiss model in the parameter regime that $B\in(0,1)$ and $J=1$. Let us start on the classical side, i.e. we consider the Poisson manifold $B^3$ with Poisson bracket defined in \eqref{pbra}. The relevant symmetry group is $\bZ_2\cong\{\pm 1\}$, also called {\em parity} or {\em mirror} symmetry. The non-trivial element $(-1)$ that implements the symmetry $\gamma:G\to\text{Aut}(C(B^3))$ is given by the automorphism
\begin{align}\label{symmetryonthreeball}
(\gamma_{-1}f)(x,y,z)=f(x,-y,-z); \ f\in C(B^3),\ (x,y,z)\in B^3.
\end{align}
The time evolution $\alpha_t^{h_0^{CW}}$ generated by the Hamiltonian flow $\phi^{h_0^{CW}}$ is clearly complete on the (compact) manifold $B^3$. It is not difficult to see that $\gamma$ commutes with $\alpha^{h_0^{CW}}$, so that $\bZ_2$ is dynamical symmetry group of $(C(B^3), \alpha^{h_0^{CW}})$.

The ground states are convex combinations of Dirac measures concentrated at points in $B^3$ which correspond to the absolute minima of $h_0^{CW}$. We have seen that these points are given by \eqref{sphericalminima}. It follows that the Dirac measures $\mu^{(0)}_\pm$ localized at $\mathbf{x}_\pm:=(B,0,\pm\sqrt{1-B^2})$, or the corresponding functionals  $\omega^{(0)}_\pm$ on $C(B^3)$, given by $\omega^{(0)}_\pm(f):=f(\mathbf{x}_\pm)$ are not invariant under the symmetry \eqref{symmetryonthreeball}: $\omega^{(0)}_\pm\circ\gamma_{-1}= \omega^{(0)}_\mp$.
Therefore, the $\bZ_2$-symmetry in the classical CW model is spontaneously broken. 
\\\\
On the quantum side instead it can be shown that for any $N<\infty$  the extremal (or in this case also the pure) algebraic vector state $\omega_{1/N}^{(0)}(\cdot):=\langle\Psi_{N}^{(0)},(\cdot)\Psi_{N}^{(0)}\rangle$  induced by the vector $\Psi_{N}^{(0)}$ is unique \cite{VGRL18}. The relevant $\bZ_2$- symmetry on $M_2(\bC)^{\otimes N}$ is locally implemented by the operator $\sigma_1$, i.e. it is given by the $N$-fold tensor of the automorphism on $M_2(\bC)$ defined by
\begin{align}\label{autosigma1}
A\mapsto\sigma_1A\sigma_1^*.
\end{align}
If $\zeta$ is the nontrivial element $(-1)$ of $\bZ_2$, we denote the automorphism of $M_2(\bC)^{\otimes N}$ induced by \eqref{autosigma1} by $\zeta_{N}$, i.e.
\begin{align}
\zeta_{N}(A)=V_NAV_N^*, \ \ (A\in M_2(\bC)^{\otimes N});
\end{align}
where $V_N:=\otimes_{n=1}^N\sigma_1$ .  The time evolution on $M_2(\bC)^{\otimes N}$ is defined by
\begin{align}
\alpha_t^{N}(A)=U_{-t}^NAU_t^N,\ \ (A\in M_2(\bC)^{\otimes N});
\end{align}
with $U_t^N=e^{-itH_{1/N}^{CW}}$. Since locally, for each $N\in\mathbb{N}$
\begin{align}
[V_N,H_{1/N}^{CW}]=0,
\end{align}
it follows that that $G=\bZ_2$ defines a dynamical symmetry group. Uniqueness of $\Psi_{N}^{(0)}$ now implies that the  state $\omega_{1/N}^{(0)}$ is strictly  invariant under $\bZ_2$, i.e.
\begin{align}
\omega_{1/N}^{(0)}\circ \zeta_{N}=\omega_{1/N}^{(0)}.
\end{align}
In addition, it is easy to see that the algebraic state $\omega_{1/N}^{(0)}$ also qualifies a ground state according to Definition \ref{def:grounds} \cite{LMV,vandeVen}.
In summary, no SSB occurs in the $C^*$-dynamical system $(M_2(\bC)^{\otimes N}, \alpha^{N})$.  The classical limit \eqref{question} predicted by Theorem \ref{mainsecond2} is a $\bZ_2$-invariant, but  mixed ground state of the $C^*$-dynamical system $(C(B^3),\alpha^{h_0^{CW}})$. Hence, at least at the level of ground states we observe that spontaneous symmetry breaking shows up as emergent phenomenon when passing from the quantum to the classical world by sending $N\to\infty$.

\subsection{Bose-Hubbard model}\label{applicationsBH}

The final example we consider is the quantum Bose-Hubbard model of $N$ bosons (with $S=0$) on a chain of size $K$ \cite{LiebBH}. For each site $i$ we denote by $a_i^\dagger$ and $a_i$ the creation and the annihilation operators, respectively, of a boson at site $i$. We define the local number operators by $n_i=a_i^\dagger a_i$ and the total number operator by $N=\sum_in_i$,  which denotes the total number of particles and is a conserved quantity. The Hamiltonian is
\begin{align}\label{BHmodel}
H_{N}^{BH}=-\sum_{\langle i,j \rangle}^Kt_{ij}a_i^\dagger a_j+\sum_{i=1}^KU_{i}(n_i-\frac{1}{2})^2-\rho\sum_{\langle i,j\rangle}^Kt_{ij}n_in_j,
\end{align}
where $T$ denotes the  hopping matrix describing the mobility of bosons, with elements $t_{ij}$, and we assume, as a convention, that $t_{ij} = 0$ if $i$ and $j$ are non nearest neighbors. Moreover, we assume $T$ to be symmetric, i.e. $t_{ij}=t_{ji}$. At each site $i$ there is also given a number $U_i$ modeling the  the on-site interaction $U_{i}$ which can be attractive ($U_i<0$) or repulsive ($U_i>0$). The notation $\langle i,j \rangle$ indicates summation over all nearest neighbors. The symbol $\rho$ indicates a nearest-neighbour interaction term. It can be shown that the dimension of the Hilbert space of this model is given by the number $D_{K,N}$ defined by 
\begin{align}
D_{K,N}={N+K-1 \choose K-1}.\label{dimension}
\end{align}
This Hilbert space can be seen as the carrier space of the rank $N$ totally symmetric irreducible representation of $SU(K)$, which in turn can be identified with the vector space $\Pi_N$ of homogeneous polynomials in $K$ complex variables of fixed degree $N$ \cite{Gitman}. A suitable basis is given in terms of the monomials 
\begin{align}\label{basis}
&\Psi_{N,\{n\}}(z)=\sqrt{\frac{N!}{n_1!\cdot\cdot\cdot n_K!}}z_1^{n_1}\cdot\cdot\cdot z_K^{n_K},\\
&\{n\}=\{n_1,...,n_K \ | \ \sum_{i=1}^Kn_i=N\}.
\end{align}
For convenience we focus on the case $K=2$, so that $D_{2,N}=N+1$, and to illustrate spontaneous symmetry breaking we fix the parameters to be $T=1$, $U=-2$ and $\rho=-2$.\footnote{Similar as to the CW-model one can take these into a small range.} Analogously to the quantum Curie-Weiss model the quantum Bose-Hubbard model admits a classical counterpart as well, but this time defined on $S^2$, which is just the familiar {\em Bloch sphere} from physics.\footnote{We stress that for general $K$, the algebra $B(\Pi_N)$ of bounded operators on $\Pi_N$  admits a classical counterpart given by the commutative $C^*$ algebra of continuous functions on the complex projective space $\mathbb{C}\mathbb{P}^{K-1}$ \cite{Gitman}. Similar as the case $K=2$ for which  the BH-model \eqref{BHmodel} has a classical analog (cf. \eqref{clasBH}) related by deformation quantization of $S^2$ (viz. quantization maps \eqref{defquan3} and Table \ref{TABLE}), for general $K$ it has a classical analog on $\mathbb{C}\mathbb{P}^{K-1}$, and these are in turn related by deformation quantization of $\mathbb{C}\mathbb{P}^{K-1}$, which is well-known \cite{BMS94}.} To see this, we consider similar as in the case of the quantum Curie-Weiss model a normalized version of this model in the following way, i.e. we replace the operators $a_i^\dagger a_j$  by $1/(N+1)a_i^\dagger a_j$, so that the normalized Bose-Hubbard Hamiltonian reads
\begin{align}
H_{N,nor}^{BH}\equiv H_{N}^{BH}/(N+1) =-\frac{1}{N+1}(a_1^\dagger a_2+a_2^\dagger a_1)-\frac{2}{(N+1)^2}\bigg{(}(n_1-\frac{1}{2})^2+(n_2-\frac{1}{2})^2-2n_1n_2\bigg{)}.\label{BHnormalized1}
\end{align}
Using the fact that any unitary irreducible representation of the Lie algebra $\mathfrak{su}(2)$ has dimension $2J+1$, we can use the famous spin operators $S_x,S_y,S_z$ on the Hilbert space $\Pi_N$, where $N=2J$,\footnote{The number $J:=N/2$ is also called the {\em spin} of the given irreducible representation. As a result, the bosonic limit $N\to\infty$ may also be interpreted as a classical limit in the spin quantum number $J$ of a single quantum spin system.} to represent the BH-Hamiltonian in terms of $S_x$, $S_y$ and $S_z$. It follows that \eqref{BHnormalized1} reads
\begin{align}
-\frac{2}{(N+1)^2}(S_z^2-N+1/2)-\frac{1}{N+1}S_x.\label{norBHnew}
\end{align}
We now recall a result originally obtained by Lieb \cite{Lieb}, namely that under the maps $Q_{1/N}'$ given by \eqref{defquan3} one has a correspondence between functions $G$ (also called upper symbol) on the sphere $S^2$ and operators $A_G$ on $\mathbb{C}^{N+1}$ such that they satisfy the relation $A_G=Q_{1/N}'(G)$. For some spin operators, the functions $G$ are determined (see Table \ref{TABLE} below).
\begin{table}[ht]
\begin{center}
\begin{tabular}
{ |p{3cm}||p{9cm}|p{3cm}|p{3cm}|}
 \hline
 $\text{Spin Operator}$ & $G(\theta,\phi)$ \\
 \hline
 $S_z$   &  $\frac{1}{2}(N+2)\cos{(\theta)}$\\
 $S_z^2$ &  $\frac{1}{4}(N+2)(N+3)\cos{(\theta)}^2-\frac{1}{4}(N+2)$\\
 $S_x$ & $\frac{1}{2}(N+2)\sin{(\theta)}\cos{(\phi)}$ \\
 $S_x^2$ & $\frac{1}{4}(N+2)(N+3)\cos{(\phi)}\sin {(\theta)}-\frac{1}{4}(N+2)$\\
 $S_y$ & $\frac{1}{2}(N+2)\sin{(\theta)}\sin{(\phi)}$ \\
 $S_y^2$ &  $\frac{1}{4}(N+2)(N+3)\sin{(\phi)}\sin{(\theta)}-\frac{1}{4}(N+2)$ \\
 \hline
\end{tabular}
\caption{Spin operators on $\text{Sym}^N(\mathbb{C}^2)\simeq\mathbb{C}^{N+1}$ and their corresponding upper symbols $G$.}
\label{TABLE}
\end{center}
\end{table}
Using these results, a straightforward computation shows that the Hamiltonian $H_{N,nor}^{BH}$ satisfies
\begin{align}
H_{N,nor}^{BH}=Q_{1/N}'(h_{0}^{BH}) +O(1/N).
\end{align}
where $O(1/N)$ is meant in operator norm and the function $h_{0}^{BH}\in C(S^2)$ in spherical coordinates $(\theta,\phi)$, is given by
\begin{align}
h_{0}^{BH}(\theta,\phi)=-\frac{1}{2}\bigg{(}\sin{(\theta)}\cos{(\phi)} +\cos^2{(\theta)}\bigg{)}, \ \ (\theta\in [0,\pi], \ \phi\in [0, 2\pi)).\label{clasBH}
\end{align}
This function is what we call the {\bf classical Bose-Hubbard} model.
\\\\
Let us now discuss the concept of SSB in this model, starting on the classical site. The relevant symmetry group acting on $S^2$ is again the group $\mathbb{Z}_2$ identified with $\{\pm 1\}$, the non-trivial element $-1$ inducing the automorphsim $\gamma_{-1}\equiv\gamma$ acts as follows on functions $f$ on the sphere $S^2$,
\begin{align}\label{actiontwospheres}
(\gamma f)(\theta,\phi)=f(\pi-\theta,-\phi), \ \ (\theta\in [0,\pi], \phi\in [0,2\pi)).
\end{align}
Also in this case, the time evolution $\alpha_t^{h_0^{BH}}$ generated by Hamiltonian flow $\phi^{h_0^{BH}}$ is complete, and  commutes with $\gamma$. In other words, $\bZ_2$ is a dynamical symmetry group of $(C(S^2), \alpha^{h_0^{BH}})$. Analogously to the Curie-Weiss model we note that unlike the case where $N$ is finite, the ground state of the classical BH hamiltonian \eqref{clasBH} is not unique.  Indeed, following Remark \ref{weakversusspon} it suffices to consider the points $\mathbf{\Omega}\equiv(\theta,\phi)$ on which the classical BH attains an absolute minima, i.e. $\{\mathbf{\Omega_-}=(\pi/6,0), \mathbf{\Omega_+}=(5\pi/6,0)\}$. By a similar argument as before we can say that the $\bZ_2$-symmetry in the classical BH model is spontaneously broken, since neither $\mathbf{\Omega_+}$ nor $\mathbf{\Omega_-}$ is invariant under this symmetry: instead, $\mathbf{\Omega_+}$ is mapped to $\mathbf{\Omega_-}$.
\\\\
Let us consider the quantum Bose-Hubbard model. To prove that the ground state is non-degenerate we may rely on the following observation. It is not difficult to see that the normalized BH Hamiltonian \eqref{norBHnew} in the basis $\{\Psi_{N,\{n_1,N-n_1\}}\}_{n_1=0,..,N}$ becomes a tridiagonal matrix of dimension $(N+1)$ with elements
\begin{align}
&H_{N,nor}^{BH}(n_1+1,n_1+1)=-\frac{2}{(N+1)^2}\bigg{(}(2n_1-N)^2+N-1/2\bigg{)}, \ (n_1=0,...,N);\\
&H_{N,nor}^{BH}(n_1+2,n_1+1)=-\frac{1}{N+1}\sqrt{(N-n_1)(n_1+1)}, \ (n_1=0,..,N-1); \\
&H_{N,nor}^{BH}(n_1,n_1+1)=-\frac{1}{N+1}\sqrt{n_1(N-n_1+1)}, \ (n_1=1,..,N).
\end{align}
Since this tridiagonal matrix is real-symmetric and all off-diagonal terms are non-zero, by basic arguments from linear algebra it follows that all eigenvalues are real and distinct. As a result, each eigenspace is one-dimensional. In particular, for each $N$ the ground state eigenvector of $H_{N,nor}^{BH}$ (and thus also of \eqref{BHmodel}) is unique, up to a constant, and by the Perron-Frobenius Theorem it can be chosen to have strictly positive components. As for the CW model we denote this eigenvector by $\Psi_{N}^{(0)}$ as well, which as a result of the choice of this basis is an element of $\Pi_N$.
\\\\
The $\bZ_2$-symmetry of the quantum model (induced by the non-trivial element of the group)  is implemented by the unitary operator $\tilde{U}_N$ defined on the monomials 
\begin{align}
\tilde{U}_{N}\Psi_{N,\{n_1,N-n_1\}}(z_1,z_2)=\Psi_{N,\{n_1,N-n_1\}}(z_2,z_1),
\end{align}
and extended by linearity to all homogeneous polynomials of degree $N$. The ensuing automorphism $\tilde{\zeta}_{N}$ on the $C^*$-algebra $B(\Pi_N)$ is defined as
\begin{align}
\tilde{\zeta}_{N}(A)=\tilde{U}_NA\tilde{U}_N^*, \ \ (A\in B(\Pi_N)),
\end{align}
and time evolution
\begin{align}
\tilde{\alpha}_t^{N}(A)=\tilde{U}_{-t}^NA\tilde{U}_t^N, \ \ (A\in B(\Pi_N)),
\end{align}
where $\tilde{U}_t^N=e^{-itH_{N,nor}^{BH}}$. Again, for $N\in\mathbb{N}$
\begin{align}
\tilde{\zeta}_{N}\circ\tilde{\alpha}_t^{N}=\tilde{\alpha}_t^{N}\circ\tilde{\zeta}_{N},
\end{align}
so that $G=\bZ_2$ defines a dynamical symmetry group. Uniqueness of $\Psi_{N}^{(0)}$ now implies that the algebraic vector state $\omega_{1/N}^{(0)}$ is strictly  invariant under $\bZ_2$: no SSB occurs for any finite $N$.  In a similar fashion as above (see for example \cite{MV} and references therein) one can prove that the algebraic ground state $\omega_{1/N}^{(0)}(\cdot)=\langle\Psi_{N}^{(0)},\cdot\Psi_{N}^{(0)}\rangle$ admits a classical limit with respect to the observables induced by the maps \eqref{defquan3}\footnote{With this expression we mean the following. Strictly  speaking, the quantization maps defined through \eqref{defquan3} are initially defined on $B(\text{Sym}^N(\bC^2))$. Since $\Pi_N\cong\text{Sym}^N(\bC^2)$ it is possible to map the operators $Q_{1/N}'(f)$ to the space $B(\Pi_N)$ by means of a unitary transformation. In this fashion, as $\Psi_{N}^{(0)}\in\Pi_N$, the expression $\langle\Psi_{N}^{(0)},Q_{1/N}'(f)\Psi_{N}^{(0)}\rangle$ indeed makes sense.} in the sense that,
\begin{align}\label{laslimBH2}
\lim_{N\to\infty}\omega_{1/N}^{(0)}(Q_{1/N}'(f))=\frac{1}{2}(f(\mathbf{\Omega_-})+f(\mathbf{\Omega_+})), \ \ (f\in C(S^2)),
\end{align}
where the limit in the number of bosonic particles $N$ may also be seen (via the relation $N=2J$) as the limit in the spin quantum number $J\to \infty$ of a single quantum spin system.
This limit is a $\bZ_2$-invariant, but mixed ground state of the $C^*$-dynamical system $(C(S^2),\alpha^{h_0^{BH}})$ which therefore breaks the $\bZ_2$-symmetry.  Again, we see that this algebraic approach allows to study SSB as emergent phenomenon: it only arises in the classical limit predicted by \eqref{laslimBH2}.

\section{Discussion}\label{Discussion}
In this paper we discussed the concept of the classical limit from an algebraic point of view. We have seen that the theory of a $C^*$-bundle and deformation quantization provide a convenient setting to study this limit from a rigorous point of view. In addition, this framework allows to discuss the natural phenomenon of spontaneous symmetry breaking (SSB) formulated in terms of $C^*$-algebras, and it confirms that SSB occurs as emergent phenomenon. These ideas have been highlighted through several physical examples of different origin.


\subsection{Symmetry breaking in Nature}\label{SSBINNATURE}

The previous examples have shown that SSB is a natural phenomenon emerging in the limit $N\to\infty$ (where $N$ plays the role of the number of lattice sites, the number of bosons or the spin quantum number) or in the context of Schr\"{o}dinger operators, in Planck's constant $\hbar\to 0$. Indeed, a pure $G$-invariant state typically converges to a mixed $G$-invariant state. However, even though this $C^*$-algebraic notion of SSB is theoretically correct it is not the whole story, since according to Definition \ref{def:SSB}, SSB also occurs whenever pure (or more generally, extreme) ground states are not invariant under the pertinent symmetry group $G$. It is precisely the latter notion of SSB that occurs in Nature: the limiting ground state is typically observed to be asymmetric in the sense that it is pure but not $G$-invariant. This state is therefore also called the ``physical'' ground state. 

Relying on the physical idea that the limiting state should be pure, but not $G$-invariant, one should introduce a quantum ``ground-ish" state that converges to a pure, but not $G$-invariant  physical classical ground states rather than to the unphysical $G$-invariant mixture predicted by theory. The mechanism to accomplish this, originating with Anderson (1952), is based on forming symmetry-breaking linear combinations of low-lying states (sometimes called “Anderson’s tower of states”) whose energy difference vanishes in the pertinent limit. In the limit (i.e. either $\hbar\to 0$  or $N\to\infty$) these low-lying eigenstates, still defining a pure state, converge to some symmetry-breaking pure ground state on the limit system (be it a classical system or an infinite quantum system). 

Even though this approach yields the right  ``physical'' ground state  in the pertinent limit, it still does not account for the fact that in Nature real and hence {\em finite} materials (such as crystals, antiferromagnets, etc.) evidently display spontaneous symmetry breaking, since in Theory it seems  forbidden in such systems (since, as we have seen, it allows SSB only in classical or infinite quantum systems.)
 The solution to this paradox is based on the ideas stemmed from Butterfield ~\cite{Butterfield}: ``emergent phenomena is behaviour that is novel and
robust relative to some comparison class''. In our situation this can be interpreted in the sense that a robust form of symmetry breaking should occur before the pertinent limit (viz. finite $N$ or non-zero $\hbar$) and it is precisely this behaviour which is physically real. This can be achieved by some form of
perturbation theory~\cite{Sim85,VGRL18}. Although in a different context, this approach, firstly introduced by Jona-Lasinio et al. \cite{Jona} and later called the “flea on the elephant” by Simon \cite{Sim85}, is based on the fact that the exact ground state of a tiny perturbed Hamiltonian approximates the right physical (symmetry-broken) state in such a way that this perturbed ground state is $G$-invariant for relatively large values of $\hbar>0$ (or similarly, for small values of $N$), but when approaching the relevant limiting regime (i.e. $\hbar<<1$ or $N>>0$) the perturbed ground state loses its $G$-invariance,\footnote{We remind the reader that for any $\hbar>0$ or $N<\infty$ the unperturbed ground state of a generic Hamiltonian is typically $G$-invariant, so that no SSB occurs. Therefore, strictly speaking any approach to symmetry breaking in Nature ($\hbar>0$ or $N<\infty$ ) is {\em explicit} rather than spontaneous. } and therefore breaks the symmetry already {\em before} the ensuing limit.\footnote{The physical intuition behind this mechanism is that these tiny perturbations should arise naturally and might correspond either to imperfections of the material or contributions to the Hamiltonian from the (usually ignored) environment.}  In other words this mechanism provides a correct interpretation of symmetry breaking in Nature, which is therefore also compatible with the physical idea that the symmetry of the system in question should become highly sensitive to small perturbations in the relevant regime (viz. Section \ref{introduction}).

It is a challenging problem to study this form of symmetry breaking occuring from an algebraic point of view. Up to my knowledge no quantization procedures yet have been investigated to prove that the classical limit associated to a sequence of symmetry breaking low-lying eigenstates, or to the exact ground state of a slightly perturbed Hamiltonian, yields a pure but non $G$-invariant and hence a physical ground state. 

\section*{Acknowledgments} The research was financially supported by the Istituto  Nazionale di Alta Matematica “Francesco Severi'' as part of the European project INdAM-DP-Cofund-2015 “INdAM Doctoral Programme in Mathematics and/or Application Cofunded by Marie Sk\l odowska-Curie Actions” under grant number: 800 713485. The author thanks Valter Moretti, Francesco Fidaleo and Nicolò Drago for their feedback.


\begin{thebibliography}{999}
\bibitem{ABN}   A.E. Allahverdyana, R. Balian, and Th. M. Nieuwenhuizen,  
Understanding quantum measurement from the solution of dynamical models,  {\em Physics Reports},
525, 1--166 (2013).
\bibitem{And} P. W. Anderson, An approximate quantum theory of the antiferromagnetic ground state, {\em Phys Rev. 86}, 694 (1952).
\bibitem{Batterman} R. Batterman. {\em The Devil in the Details: Asymptotic Reasoning in Explanation, Reduction,
and Emergence.} Oxord University Press (2002).
\bibitem{Bay} Bayen, F., Flato, M., Fronsdal, C., Lichnerowicz, A., Sternheimer, D.  Deformation theory and quantization I, II. {\em Annals of Physics}, 111, 61--110, 111--151 (1978).
\bibitem{Ber} F.A. Berezin, General concept of quantization, {\em Communications in Mathematical Physics}, 40, 153--174 (1975).
\bibitem{Berry} M.V. Berry, Singular limits, {\em Physics Today}, 55, 10--11 (2002).
\bibitem{Bona} P. Bona, The dynamics of a class of mean-field theories. {\em Journal of Mathematical Physics} 29, 2223--2235 (1988).
\bibitem{BMS94} M. Bordemann, E. Meinrenken, and M. Schlichenmaier, Toeplitz quantization of K\"{a}hler manifolds and
$gl(N)$, $N\to\infty$ limits,  {\em  Communications in Mathematical Physics} 165, 281--296 (1994).  
\bibitem{BRW} M. Bordermann, H. R\"{o}mer, S. Waldmann, A Remark on Formal KMS States in Deformation Quantization,  {\em Letters in Mathematical Physics}, 45, 49--61 (1998).
\bibitem{BR1} O. Bratteli and  D.W. Robinson, {\em Operator Algebras and Quantum Statistical Mechanics. Vol.\ I:
Equilibrium States, Models in Statistical Mechanics.}, Springer (1981). 
\bibitem{BR2} O. Bratteli and  D.W. Robinson, {\em Operator Algebras and Quantum Statistical Mechanics. Vol.\ II:
Equilibrium States, Models in Statistical Mechanics.}, Springer (1981). 
\bibitem{Butterfield} J. Butterfield, Less is different: Emergence and reduction reconciled, {\em Foundations of
Physics}, 41, 1065–1135 (2011).
\bibitem{CCIL} L. Chayes, N. Crawford, D. Ioffe, and A. Levit, The phase diagram of the quantum Curie--Weiss model, {\em Journal of Statistical Physics}, 133, 131--149  (2008).
\bibitem{Dix} J. Dixmier, {\em C*-algebras}, North-Holland (1977). 
\bibitem{DW} N.G. Duffield and R.F.  Werner,  Local dynamics of mean-field quantum systems, {\em Helvetica Physica Acta}, 65, 1016--1054 (1992). 
\bibitem{FV2017} S. Friedli and Y Velenik, {\em Statistical Mechanics of Lattice Systems: A Concrete Mathematical Introduction}, Cambridge University Press (2017). 
\bibitem{GV} G. Gallavotti, E. Verboven, On the classical KMS boundary condition, {\em Il Nuovo Cimento B}, 28 N. 1, 274--286 (1975).
\bibitem{Gitman} D. M. Gitman, A. L. Selepin,  Coherent states of $SU(N)$ groups. {\em Journal of Physics A: Mathematical and General}, 26, (1993).
\bibitem{GriSjo} A. Grigis, J. Sj\"{o}strand, {\em Microlocal Analysis for Differential Operators, an Introduction}, London Mathematical Society Lecture Note Series (2013).
\bibitem{Hel} Helffer, B. {\em Semi-classical Analysis for the Schr\"{o}dinger Operator and Applications.}, Heidelberg: Springer (1988).
\bibitem{SJHE} B. Helffer, J. Sj\"{o}strand. Multiple Wells in the Semi-Classical Limit 1, {\em Communications in Partial Differential Equations}, 9, 337--408 (1984).
\bibitem{Jona}  G. Jona-Lasinio, F. Martinelli and E. Scoppola,  New approach to the semiclassical limit of quantum mechanics, {\em Communications in Mathematical Physics}, 80, 223 (1981).
\bibitem{Kirchberg-Wassermann} E. Kirchberg, S. Wassermann, Operations on continuous bundles of $C^*$algebras. {\em Mathematische Annalen}, 303, 677--697 (1995).
\bibitem{KoTa1993} T. Koma and H. Tasaki,  Symmetry breaking in Heisenberg Antiferromagnets, {\em Communications in Mathematical  Physics}, 158, 191-214 (1993).
\bibitem{KoTa} T. Koma and H. Tasaki,  Symmetry breaking and finite-size effects in quantum many-body systems, {\em Journal of Statistical Physics}, 76, 745--803 (1994).
\bibitem{Lan98} N.P. Landsman, {\em Mathematical Topics Between Classical and Quantum Theory}, Springer (1998).
\bibitem{Lan13} N.P. Landsman, Spontaneous Symmetry Breaking in Quantum Systems: Emergence or Reduction? {\em Studies in History and Philosophy of Science Part B: Studies in History and Philosophy of Modern Physics}, 44, 379--394 (2013).
\bibitem{Lan17} K. Landsman, {\em Foundations of Quantum Theory: From Classical Concepts to Operator Algebras}, Springer (2017). Open Access at  \verb#http://www.springer.com/gp/book/9783319517766#.
\bibitem{LMV} K. Landsman, V.Moretti, C.J.F. van de Ven,  Strict deformation quantization of the state space of $M_k(\bC)$ with applications to the Curie-Weiss model.
{\em Reviews in Mathematical Physics}, 32 (2020).
\bibitem{Lieb} E.H. Lieb, The classical limit of quantum spin systems. {\em Communications in Mathematical Physics}, 62, 327--340 (1973). 
\bibitem{LiebBH} E.H. Lieb, The Hubbard model: some rigorous results and open problems. In {\em Condensed Matter Physics and Exactly Soluble Models}, 59--77, Springer (2004).
\bibitem{Mor} V. Moretti, {\em Spectral Theory and Quantum Mechanics}, 2nd Edition, Springer (2018).
\bibitem{MV} V. Moretti, C.J.F. van de Ven,  Bulk-boundary asymptotic equivalence of two strict deformation quantizations, {\em Letters in Mathematical Physics}, 110,  2941--2963 (2020).
\bibitem{MorVen2} V. Moretti, C.J.F. van de Ven, The classical limit of Schr\"{o}dinger operators in the framework of Berezin quantization and spontaneous symmetry breaking as emergent phenomenon. {\em Int. J. Geom. Methods Mod. Phys.} 19, No. 01, 2250003 (2022).
\bibitem{MV2} S. Murro, C.J.F. van de Ven, Injective tensor products in strict deformation quantization. {\em Math. Phys. Anal. Geom.} 25, No. 2 (2022).
\bibitem{Pe72} A.M. Perelomov, Coherent states for arbitrary Lie groups, {\em  Communications in Mathematical Physics}, 26, 222--236 (1972).
\bibitem{RW} G.A. Raggio and R.F. Werner, Quantum statistical mechanics of general mean field systems, {\em  Helvetica Physica Acta}, 62,  980--1003 (1989).
\bibitem{RS2} M. Reed and B. Simon, {\em Methods of Modern Mathematical Physics}, Vol II, Academic Press (1975).
\bibitem{RS4} M. Reed and B. Simon, {\em Methods of Modern Mathematical Physics}, Vol IV, Academic Press (1975).
\bibitem{Rie89} M.A. Rieffel, Deformation quantization of Heisenberg manifolds, {\em  Communications in Mathematical Physics}, 121, 531--562 (1989).
\bibitem{Rie94} M.A. Rieffel, Quantization and $C^*$-algebras, {\em Contemporary Mathematics}, 167, 67--97 (1994). 
\bibitem{Rudin} W. Rudin, {\em Real and complex analysis}, McGraw-Hill  (1986).
\bibitem{Ruelle} D. Ruelle, Natural nonequilibrium states in quantum statistical mechanics, {\em Journal of Statistical Physics}, 98, 57--75 (2000).
\bibitem{Sim85} B. Simon, Semiclassical analysis of low lying eigenvalues. IV. The flea on the elephant. {\em Journal of Functional Analysis}, 63, 123 (1985).
\bibitem{Tasaki19} H. Tasaki, Long-Range Order, “Tower” of States, and Symmetry Breaking in Lattice Quantum Systems. {\em Journal of Statistical Physics}, 174 (2019).
\bibitem{vandeVen} C. J. F. van de Ven,  {\em Properties of Quantum Spin Systems and their Classical Limit}
(M.Sc. Thesis, Radboud University, 2018), \verb#https://www.math.ru.nl/~landsman/Chris2018.pdf#.
\bibitem{Ven2020} C.J.F. van de Ven,  The classical limit of mean-field quantum theories. {\em Journal of Mathematical Physics}, 61, 121901 (2020).
\bibitem{VGRL18}  C. J. F. van de Ven, G. C. Groenenboom, R. Reuvers, N. P. Landsman,  Quantum spin systems versus Schr\"{o}dinger operators: A case study in spontaneous symmetry breaking {\em SciPost}, 8, 022, (2020). 
\bibitem{Wez1} J. van Wezel,  Quantum dynamics in the thermodynamic limit, {\em  Phys. Rev. B}, 78, 054301 (2008).
\bibitem{Wez2} J. van Wezel and J. van den Brink,  Spontaneous symmetry breaking in quantum mechanics, {\em American Journal of Physics}, 75, 635 (2007).
\bibitem{MZ} M. Zworski {\em Semiclassical Analysis}, Graduate Studies in Mathematics 138, American Mathematical Society (2012).
\end{thebibliography}
\end{document}